\documentclass[a4paper,12pt]{article}

\usepackage{float}
\usepackage{cite}
\usepackage[margin=3cm, includefoot, includehead]{geometry}

\usepackage{amsmath}
\usepackage{amssymb}
\usepackage[dvipdfmx]{graphicx}

\begin{document}
 
\begin{flushright}
IU-TH-9
\end{flushright}
\vskip 0.5 truecm
 
 
\begin{center}
{\Large{\bf Momentum and Hamiltonian in Complex Action Theory}}\\
\vskip 1.5cm

{\large Keiichi Nagao$^{a,b}$\footnote{e-mail
address: keiichi.nagao.phys@vc.ibaraki.ac.jp} 
and Holger Bech Nielsen$^b$\footnote{e-mail
address: hbech@nbi.dk}}
\vskip 0.5cm

$^a${\it Faculty of Education, Ibaraki University, Bunkyo 2-1-1, Mito 310-8512 
Japan  }\\
$^b${\it Niels Bohr Institute, University of Copenhagen, 
Blegdamsvej 17, 2100 Copenhagen $\O$, Denmark}\\

\end{center}

\vskip 1cm
\begin{center}
\begin{bf}
Abstract
\end{bf}
\end{center}
In the complex action theory (CAT) 
we explicitly examine how the momentum and Hamiltonian are defined 
from the Feynman path integral (FPI) point of view based on 
the complex coordinate formalism of our foregoing paper. 
After reviewing the formalism briefly, 
we describe in FPI with a Lagrangian the time development of a $\xi$-parametrized wave function, 
which is a solution to an eigenvalue problem of a momentum operator. 
Solving this eigenvalue problem, we derive the momentum, Hamiltonian, and Schr\"{o}dinger equation. 
Oppositely, starting from the Hamiltonian we derive the Lagrangian in FPI, 
and we are led to the momentum relation again via the saddle point for $p$.
This study confirms that the momentum and Hamiltonian in the CAT 
have the same forms as those in the real action theory.  
We also show the third derivation of the momentum relation via the saddle point for $q$.

\newpage
\setcounter{footnote}{0}
\section{Introduction}
\setcounter{equation}{0}

Quantum theory is described via Feynman path integral (FPI).  
The integrand in FPI has the form of $\exp(\frac{i}{\hbar}S)$, 
where $i$ is the imaginary unit and $S$ is the action. 
The action is usually taken to be real. 
However, if we assume that the integrand is more fundamental
than the action in quantum theory, then 
we can consider a theory where the action is taken to be complex, 
by speculating that since the integrand is complex, the action can be also complex. 
Based on this assumption and other related works 
in some backward causation developments inspired by general 
relativity\cite{Nielsen:2006vc,Nielsen:2006th,Nielsen:2006td,Nielsen:2008zz} 
and the non-locality explanation of 
fine-tuning problems\cite{Bennett:1995ag, Bennett:1994yx, Bennett_talk, Bennett:1996hx}, 
the complex action theory (CAT) 
has been studied intensively by one of the authors (H.B.N) and 
Ninomiya\cite{Bled2006, Nielsen:2008cm, Nielsen:2007ak, Nielsen:2005ub}. 
Compared to the usual real action theory (RAT), the imaginary part of the action
is thought to give some falsifiable predictions. 
Indeed, many interesting suggestions have been made for Higgs mass\cite{Nielsen:2007mj}, 
quantum mechanical philosophy\cite{Nielsen:2009wb,Nielsen:2010zz}, 
some fine-tuning problems\cite{Nielsen2010qq,degenerate}, 
black holes\cite{Nielsen2009hq}, 
De Broglie-Bohm particle and a cut-off in loop diagrams\cite{Bled2010B}. 
In Ref.~\cite{Nagao:2010xu} we have studied 
the time-development of some state by a non-hermitian diagonalizable bounded Hamiltonian $H$, 
and shown that we can effectively obtain a hermitian Hamiltonian after a long time 
development by introducing 
a proper inner product\footnote{A similar inner product was studied also in Ref.~\cite{Geyer}.}. 
If the hermitian Hamiltonian is given in a local form, 
a conserved probability current density can be constructed with two kinds of wave functions. 
We note that the non-hermitian Hamiltonian is a generic one, so it does not belong to 
a class of the PT symmetric non-hermitian Hamiltonians, 
which has been intensively studied recently\cite{Geyer, Bender:1998ke, Bender:1998gh, Dorey:2001uw, Dorey:2001hi, Dorey:2007zx, Bender:2002vv, FariaFring, GeyerHeissZnojil, BenderBrody, JonesSmith:2009wx, JonesSmith:2009wy, Bender:2009jg, AndrianovCannataSokolov}. 
For reviews see Refs.~\cite{Bender:2005tb, Bender:2007nj, Mostafazadeh:2008pw, Mostafazadeh:2010yx}. 
In Ref.~\cite{Nagao:2011za}, 
we have proposed the replacement of hermitian operators of coordinate and momentum 
$\hat{q}$ and $\hat{p}$ and their eigenstates $\langle  q |$ and $\langle p |$ 
with non-hermitian operators $\hat{q}_{new}$ and 
$\hat{p}_{new}$, and ${}_m\langle_{new}~ q |$ and ${}_m\langle_{new}~ p |$ with 
complex eigenvalues $q$ and $p$, 
so that we can deal with complex $q$ and $p$ obtained at the saddle point. 
Introducing a philosophy to keep the analyticity 
in parameter variables of FPI and defining a modified set of complex conjugate, 
real and imaginary parts, hermitian conjugates and bras, 
we have explicitly constructed them for complex $q$ and $p$ 
by squeezing the coherent states of harmonic oscillators. 
In addition, extending the study of Ref.~\cite{Nagao:2010xu} to the complex coordinate formalism, 
we have investigated a system defined by a diagonalizable non-hermitian bounded Hamiltonian, 
and shown that a hermitian Hamiltonian is effectively obtained after a long time development 
also in the complex coordinate formalism.


In addition, as other works related to complex saddle point paths, 
in Refs.~\cite{Garcia:1996np} and \cite{Guralnik:2007rx} 
the complete set of solutions of the differential equations 
following from the Schwinger action principle has been obtained by generalizing 
the path integral to include sums over various inequivalent contours of integration in 
the complex plane. 
In Ref.~\cite{Pehlevan:2007eq} complex Langevin equations have been studied. 
In Refs.~\cite{Ferrante:2008yq} and \cite{Ferrante:2009gk} 
a method to examine the complexified solution set  has been investigated.  

The CAT has been studied intensively as mentioned above, 
but there remain many things to be studied. 
For instance, in the above studies it has been supposed and taken as a matter of course 
that the momentum and Hamiltonian in the CAT 
have the same forms as those in the RAT, and it has not been examined explicitly so far. 
In the RAT, we first write the lagrangian $L(q,\dot{q})$ 
in terms of a coordinate $q$ and its time derivative $\dot{q}$, and then  
define the momentum $p$ by the relation $p=\frac{\partial L}{\partial \dot{q}}$. 
Next, we define the Hamiltonian $H(q,p)$ via the Legendre transformation as 
$H(q,p)= p \dot{q} - L(q, \dot{q})$. 
By replacing $q$ and $p$ with their corresponding operators 
in the classical Hamiltonian $H$, we obtain the quantum Hamiltonian. 
This is a well-known story in the RAT, but what happens in the CAT?
Once we allow the action to be complex, various quantities known 
in the RAT can drastically change, and we can encounter various exotic situations. 
Hence we have to be careful about them.


Indeed, the momentum relation $p=\frac{\partial L}{\partial \dot{q}}$ in the CAT 
is not so trivial. 
We write $\dot{q}(t)$ as $\frac{q(t+dt)-q(t)}{dt}$. 
If we quantum mechanically have in mind that our formalism corresponds to 
the deformation of the contour of originally real $q$ and $p$, 
then using functional integral FPI and imagining to choose the $q(t)$-integrals along the real axis 
we would intuitively think that both $q(t)$ and $q(t+dt)$ could be chosen real and thus 
$\dot{q}$ would be real. 
This is seemingly a discrepancy for the equation $p=m\dot{q}$ with $m$ complex, 
and also for its naive operator interpretation by expecting that the operators 
corresponding to these dynamical variables are still hermitian. 
In fact this seeming discrepancy partly motivated us to 
construct the complex coordinate formalism in Ref.~\cite{Nagao:2011za}.  
It is the main subject of the present paper to formulate in such a way that we can 
quantum mechanically assign at least some meaning to the relation $p=\frac{\partial L}{\partial \dot{q}}$. 
Normally in quantum mechanics $q(t+dt)$ has no definite value. 
Rather in FPI the system at time $t+dt$ is given in a superposition of essentially 
all $q(t+dt)$-values. 
To see whether the momentum relation holds in the CAT or not, 
we analyze it explicitly by writing it in the form of the operator 
acting on appropriate states with a number 
$\xi$ put as replacement for $q(t+dt)$. See Eq.(\ref{phat_p_xi2}). 
We shall show in this paper that the relation $p=\frac{\partial L}{\partial \dot{q}}$ is true 
in the classical sense along the time development of 
the saddle points in the integral $dq_t$ in ${\cal D} q$. 
Indeed we put forward this saddle point interpretation of $p=\frac{\partial L}{\partial \dot{q}}$ 
as Eq.(\ref{pjmqj}). 
Since with complex mass $m$ the saddle point or classical path will typically 
be complex, there is nothing strange in $p=\frac{\partial L}{\partial \dot{q}}$ of Eq.(\ref{pjmqj})  
conceived as a saddle point relation.

%

From the above point of view, in this paper we explicitly examine  
how the momentum and Hamiltonian are 
defined in the CAT based on 
the complex coordinate formalism of Ref.~\cite{Nagao:2011za}. 
We replace hermitian operators of coordinate and momentum 
$\hat{q}$ and $\hat{p}$ and their eigenstates $\langle  q |$ and $\langle p |$ 
with non-hermitian operators $\hat{q}_{new}$ and 
$\hat{p}_{new}$, and ${}_m\langle_{new}~ q |$ and ${}_m\langle_{new}~ p |$ 
with complex eigenvalues $q$ and $p$, 
and utilize the new devices such as a modified set of complex conjugate, 
hermitian conjugates and bras to realize the philosophy of keeping the analyticity in 
parameter variables of FPI. 
In the usual way of deriving and understanding the functional integral, 
a functional integral corresponding to finite time interval -- say $[t_i, t_f]$ -- 
with endpoints fixed to specific $q(t_i)$ and $q(t_f)$ values is expressed as 
$\langle q(t_f) | e^{-iH(t_f - t_i )} | q(t_i) \rangle = \int_{-\infty}^\infty e^{- \frac{i}{\hbar} \int_{t_i}^{t_f} L dt } {\cal D}q $,  
where ${\cal D}q$ is a priori taken over real $q_t$'s. 
We claim that this interpretation is possible even for complex $q(t_f)$ and $q(t_i)$ 
by following the formalism of Ref.~\cite{Nagao:2011za}. 
We can deform formally the infinitely many contours 
as long as the action $S$ is an analytical function of all the $q_t$'s. 
In this way going to complex contours only with restriction to go from $-\infty$ to $\infty$ 
is rather trivial. 
We, however, stress that after the deformation of the contour 
-- in a different way for each moment $t$ -- the contours still start at $-\infty$ and ends at $+\infty$, 
but in between of course they are usually complex. 
That is to say, appropriate contours are chosen at each time $t$. 
The present paper has also a role of showing an explicit application of the 
formalism of Ref.~\cite{Nagao:2011za}. 
Starting from the Lagrangian $L$ in FPI, we derive the momentum relation 
by considering an eigenvalue problem of 
the momentum operator, which includes a parameter $\xi$. 
We attempt to split up a wave function $\langle q(t) | \psi \rangle$ 
into various $\xi$-components, and investigate the time-development 
of the $\xi$-parametrized wave function in FPI. 
This study elucidates that the momentum in the CAT has the same form as that in the RAT. 
In addition we also derive the Hamiltonian $H$ and Schr\"{o}dinger equation 
in FPI starting from the Lagrangian $L$ 
in the CAT. 
Next in an opposite way 
we try to check whether we can reproduce the Lagrangian starting from the 
Hamiltonian, and show that the Lagrangian is derived from the Hamiltonian.
This result confirms that the Hamiltonian in the CAT 
has the same form as that in the RAT.  
It leads us to the momentum relation again via the saddle point for $p$. 
We also give the third derivation of the momentum relation via the saddle point for $q$.

This paper is organized as follows. 
In section 2 we briefly review the complex coordinate formalism of Ref.~\cite{Nagao:2011za}. 
Introducing the philosophy of keeping the analyticity in parameter 
variables of FPI and the new devices to realize it such as 
a modified set of bra, complex conjugates, hermitian conjugate and hermiticity,  
we construct non-hermitian operators of coordinate and momentum $\hat{q}_{new}$ and 
$\hat{p}_{new}$, and the hermitian conjugates of 
their eigenstates $| q \rangle_{new}$ and $| p \rangle_{new}$ with complex eigenvalues 
$q$ and $p$. 
In section 3 we derive the momentum relation by solving an eigenvalue problem 
of the momentum operator and considering the time-development of 
the wave function solution in FPI. 
We also derive the free Hamiltonian and Schr\"{o}dinger equation 
starting from the free Lagrangian in FPI. 
Next starting from the Lagrangian with a potential term 
in FPI, we derive the momentum relation, Schr\"{o}dinger equation, 
and Hamiltonian with the potential term. 
In section 4 oppositely by using the Hamiltonian $H$ and Schr\"{o}dinger equation 
we derive the Lagrangian $L$ in FPI. 
Also, we again obtain the momentum relation via the saddle point for $p$. 
This study reveals that the momentum and Hamiltonian in the CAT 
have the same forms as those in the RAT. 
Section 5 is devoted to conclusion and outlook. 
In appendix~\ref{via_saddle_q} we explain the third derivation of the momentum relation 
via the saddle point for $q$.

\section{A brief review of the complex coordinate formalism}\label{fundamental}

In Ref.~\cite{Nagao:2011za} we have proposed the replacement of hermitian operators of coordinate and momentum 
$\hat{q}$ and $\hat{p}$ and their eigenstates $\langle q |$ and $\langle p |$ 
with non-hermitian operators $\hat{q}_{new}$ and 
$\hat{p}_{new}$, and ${}_m\langle_{new}~ q |$ and ${}_m\langle_{new}~ p |$ with 
complex eigenvalues $q$ and $p$, 
so that we can deal with complex $q$ and $p$ obtained at the saddle point. 
Introducing the philosophy of keeping the analyticity in parameter variables of FPI, 
we have defined 
the new devices such as a modified set of complex conjugate, 
real and imaginary parts, hermitian conjugates and bras, 
by means of which we have explicitly constructed 
$\hat{q}_{new}$, $\hat{p}_{new}$, and the hermitian conjugates of their eigenstates 
$| q \rangle_{new}$ and $| p \rangle_{new}$ with complex $q$ and $p$ 
by utilizing the coherent states of harmonic oscillators. 
In this section we briefly review the complex coordinate formalism of Ref.~\cite{Nagao:2011za}. 
We begin with the delta function of a complex parameter.

\subsection{The delta function}

\begin{figure}[htb]
\begin{center}
\includegraphics[height=10cm]{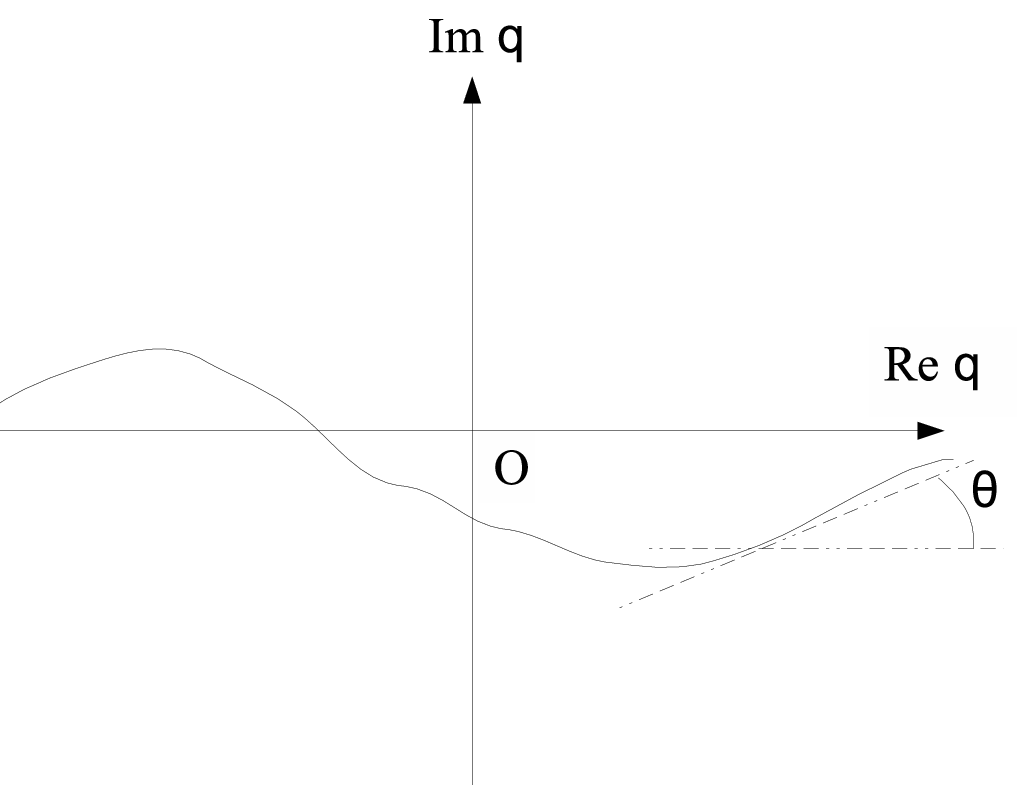}
\end{center}
\caption{An example of permitted paths}
\label{fig:contour}
\end{figure}

\begin{figure}[htb]
\begin{center}
\includegraphics[height=10cm]{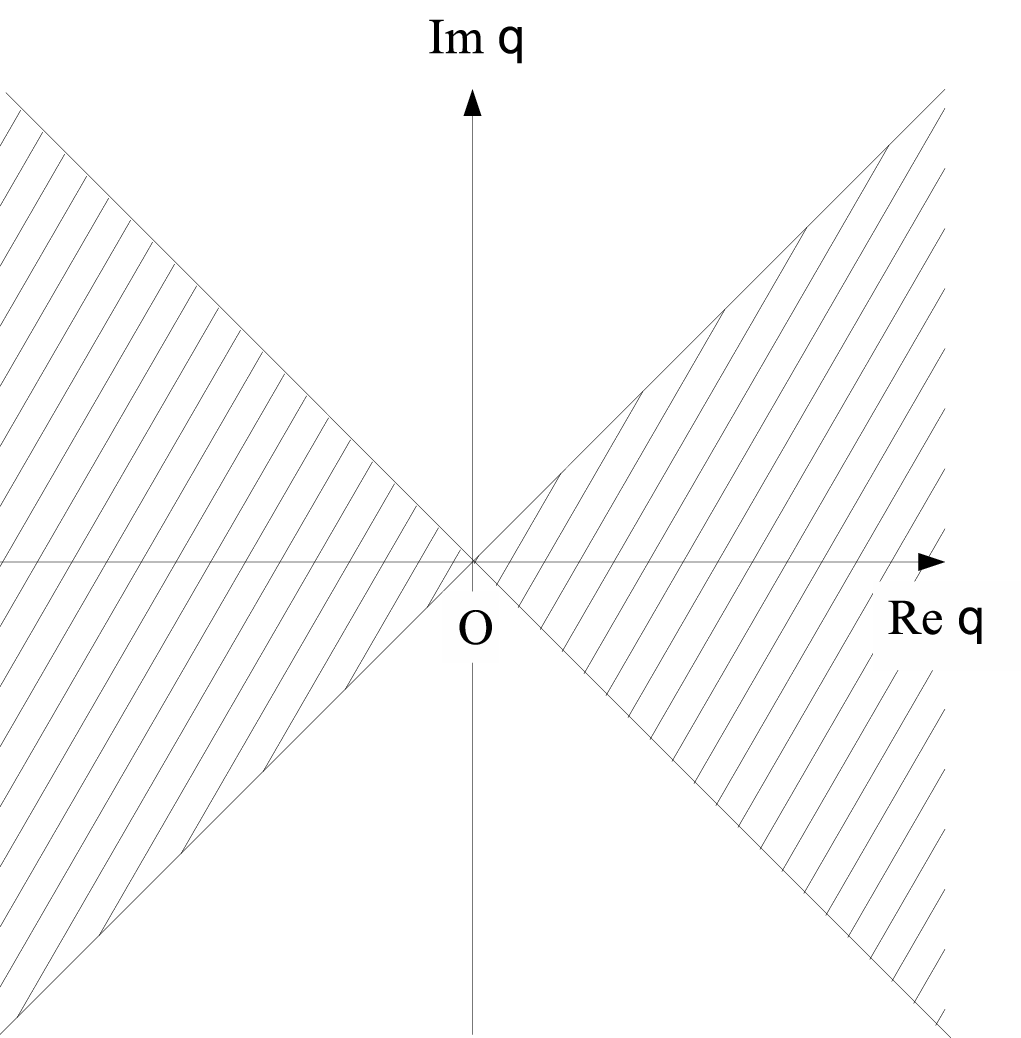}
\end{center}
\caption{Domain of the delta function}
\label{fig:delta_function}
\end{figure}

In quantum mechanics the delta function is one of essential tools 
in a theory which has orthonormal bases with continuous parameters, 
and the parameters are usually real in the RAT. 
On the other hand, parameters in the CAT can be complex in general, 
so we need to define the delta function of complex parameters. 
The delta function for real $q$ is represented as 
\begin{eqnarray}
\delta(q)&=& 
\frac{1}{2\pi} \int_{-\infty}^\infty  e^{ikq} dk ,
\end{eqnarray}
where $k$ is real. 
We assume that $k$ is also complex but asymptotic value of $k$ is real. 
In this case we can take an arbitrary path running from $-\infty$ to $\infty$ in the complex plane. 
We call such a path $C$ 
and define $\delta_c(q)$ for complex $q$ by  
\begin{equation}
\delta_c(q) \equiv \lim_{\epsilon \rightarrow +0}  \delta_c^\epsilon(q) , 
\end{equation}
where 
\begin{eqnarray}
\delta_c^\epsilon(q) &\equiv& 
\frac{1}{2\pi} \int_C e^{ikq - \epsilon k^2} dk \nonumber \\
&=& \sqrt{\frac{1}{4 \pi \epsilon}} e^{-\frac{q^2}{4\epsilon}} .
\label{delta_c_epsilon(q)}
\end{eqnarray}
In the first line of Eq.(\ref{delta_c_epsilon(q)}) we have introduced 
a finite small positive real number $\epsilon$, and in the second equality 
we have assumed that $|k|$ goes larger than $\frac{1}{\sqrt{\epsilon}}$. 
Eq.(\ref{delta_c_epsilon(q)}) is convergent for complex $q$ such that 
\begin{equation}
\left( \text{Re}(q) \right)^2 > \left( \text{Im}(q) \right)^2 \label{cond_of_q_for_delta}, 
\end{equation}
Since for any analytical test function $f(q)$\footnote{Due to the Liouville theorem if $f$ is 
a bounded entire function, 
$f$ is constant. So we are considering as $f$ an unbounded entire function or a function 
which is not entire but is holomorphic at least in the region on which the path runs.} 
the path $C$ of the integral 
$\int_C f(q) e^{-\frac{q^2}{4\epsilon}} dq $ is independent of finite $\epsilon$, 
this $\delta_c(q)$ satisfies for any $f(q)$ 
\begin{equation}
\int_C f(q) \delta_c(q) dq = f(0) ,
\end{equation}
as long as we choose a path such that at any $q$ 
its tangent line and a horizontal line 
form an angle $\theta$ whose absolute value  
is within $\frac{\pi}{4}$ to satisfy the inequality (\ref{cond_of_q_for_delta}). 
An example of permitted paths is shown in fig.\ref{fig:contour}, 
and the domain of the delta function is drawn in fig.\ref{fig:delta_function}. 
At the origin $\delta_c(q)$ is divergent. 
In the domain except for the origin, which is painted with inclined lines, 
$\delta_c(q)$ takes a vanishing value, 
while in the blank region the delta function is oscillating and divergent. 
It is well-defined for $q$ such that the condition 
(\ref{cond_of_q_for_delta}) is satisfied. 
%

\subsection{The philosophy and the new devices to handle complex parameters}\label{newdevices}

To circumvent some difficulties due to the naive extension of $q$ to complex numbers, 
which are explained in Ref.~\cite{Nagao:2011za}, 
we introduce a philosophy of keeping the analyticity in parameter variables of FPI. 
If we keep the analyticity in parameter variables of FPI, 
then we can deform a path, along which the integration is performed. 
To realize this philosophy we define a modified set of complex conjugate, 
hermitian conjugates and bras in this subsection.

\subsubsection{Modified complex conjugate $*_{ \{ \} }$}

We define a modified complex conjugate for a function of $n$-parameters 
$f( \{a_i \}_{i=1, \ldots, n} )$ by 
\begin{equation}
f(\{a_i \}_{i=1, \ldots, n} )^{*_{\{a_i | i \in A \}} } = f^*( \{a_i \}_{i \in A}  ,  \{a_i ^*\}_{i \not\in A} ) , 
\end{equation}
where $A$ denotes the set of indices attached with the parameters 
in which we keep the analyticity, 
and on the right-hand side $*$ on $f$ acts on the coefficients included in $f$. 
This is in contrast to a usual complex conjugate defined by 
\begin{equation}
f(\{a_i \}_{i=1, \ldots, n} )^{*} = f^*( \{a_i^* \}_{i=1, \ldots, n} ) .
\end{equation}
We show a set of examples of the complex conjugates $*$, $*_q$, $*_p$ and $*_{q,p}$ 
on a function $f(q,p)=a q^2 + b p^2$ as follows, 
\begin{eqnarray}
&&f(q,p)^* = f^*(q^*, p^*) = a^* (q^*)^2 + b^* (p^*)^2 , \\ 
&&f(q,p)^{*_q} = f^*(q, p^*) = a^* q^2 + b^* (p^*)^2 , \\
&&f(q,p)^{*_p} =  f^*(q^*, p) = a^* (q^*)^2 + b^* p^2 , \\
&&f(q,p)^{*_{q,p}} = f^*(q, p) = a^* q^2 + b^* p^2 ,
\end{eqnarray}
where in the first, second, third and fourth relations the analyticity is kept in no parameters, $q$, $p$ 
and both $q$ and $p$, respectively. 
For simplicity we express the modified complex conjugate as $*_{ \{  \} }$.

\subsubsection{Modified bras ${}_m \langle ~|$ and ${}_{ \{ \} } \langle ~|$, 
and modified hermitian conjugate $\dag_{ \{ \} }$ }

For some state $| \lambda \rangle$ with some complex parameter $\lambda$, 
we define a modified bra ${}_m\langle \lambda |$ by 
\begin{equation} 
{}_m\langle \lambda | = \langle \lambda^* | . \label{modified_bra_anti-linear}
\end{equation}
The modified bra is analytically extended bra with regard to the parameter $\lambda$. 
It preserves the analyticity in $\lambda$. 
In the special case of $\lambda$ being real it would become a usual bra.

Next we define a little bit generalized modified bra 
${}_{\{\}}\langle ~|$ and a modified hermitian conjugate $\dag_{ \{ \} }$ of a ket, 
where $\{ \}$ is a symbolical expression of a set of parameters 
in which we keep the analyticity. 
We show some examples, 
\begin{eqnarray}
&&{}_v \langle u | = \langle u | , \\
&&{}_{u,v}\langle u | = {}_u \langle u | = {}_m\langle u | , \label{mod_bra2} \\
&&( | u \rangle )^{\dag_{v}}  = ( | u \rangle )^\dag  = \langle u | , \\
&&( | u \rangle )^{\dag_{u, v}}  =( | u \rangle )^{\dag_{u}}  = {}_m \langle u | . 
\end{eqnarray}
We express the hermitian conjugate $\dag_{ \{ \} }$ of a ket symbolically as 
\begin{equation}
( |  ~\rangle )^{\dag_{\{ \} }} = {}_{\{ \}}\langle  ~| . 
\end{equation}
Also, we see that the hermitian conjugate $\dag_{ \{ \} }$ of a bra can be defined likewise. 
We express it as 
\begin{equation}
( {}_{\{ \}}\langle  ~| )^{\dag_{\{ \} }} =  |  ~\rangle . 
\end{equation}

For $| u \rangle$, $| v \rangle$ and some operator $A$, 
we have the following relations, 
\begin{eqnarray} 
&& \langle u | A | v \rangle^{*} = \langle v | A^{\dag}| u \rangle, 
\label{hermitian_conjugate_RAT}\\
&& {}_m\langle u | A | v \rangle^{*_{u}} = \langle v | A^{\dag} | u \rangle, 
\label{hermitian_conjugate_u}\\
&& \langle u | A | v \rangle^{*_{v }} = {}_m \langle v | A^{\dag} | u \rangle, 
\label{hermitian_conjugate_v} \\
&& {}_m\langle u | A | v \rangle^{*_{u, v }} = {}_m \langle v | A^{\dag} | u \rangle . 
\label{hermitian_conjugate_uv}  
\end{eqnarray}
In Eqs.(\ref{hermitian_conjugate_RAT})--(\ref{hermitian_conjugate_uv}) 
the analyticity is kept in no parameters, the parameter $u$, $v$, and both $u$ and $v$, respectively. 
We can choose one of the four options according to which parameters 
we keep the analyticity in. 
We can express the four relations simply as 
\begin{equation}
{}_{\{\}}\langle u | A | v \rangle^{*_{ \{ \} }} = 
{}_{ \{ \} } \langle v | A^\dag | u \rangle . \label{hermitian_conjugate_5}
\end{equation}
Various examples of the usage of the modified complex conjugate, modified bra and modified hermitian conjugate 
are given in Ref.~\cite{Nagao:2011za}.

\subsection{Non-hermitian operators $\hat{q}_{new}$ and $\hat{p}_{new}$, 
and the hermitian conjugates of their eigenstates $|q \rangle_{new}$ and $|p \rangle_{new}$}

In this subsection following Ref.~\cite{Nagao:2011za}. we summarize 
how we construct the non-hermitian operators $\hat{q}_{new}$ and $\hat{p}_{new}$, 
and their eigenstates ${}_m \langle_{new}~ q |$ and ${}_m \langle_{new}~ p |$ 
with complex eigenvalues $q$ and $p$, which satisfy the following relations, 
\begin{eqnarray}
&&{}_m \langle_{new}~ q | \hat{q}_{new} = {}_m \langle_{new}~ q | q , \label{qhatqket=qqket_newrev} \\
&&{}_m \langle_{new}~ p | \hat{p}_{new} = {}_m \langle_{new}~ p | p , \label{phatpket=ppket_newrev} \\ 
&&[\hat{q}_{new}, \hat{p}_{new} ] = i \hbar , \label{commutator_q_p_new} 
\end{eqnarray}
by utilizing coherent states of harmonic oscillators. 
Here ${}_m\langle_{new}~  q | $ and ${}_m\langle_{new}~ p | $ are modified bras of 
$| q \rangle_{new}$ and $| p \rangle_{new}$, so 
Eqs.(\ref{qhatqket=qqket_newrev}) and (\ref{phatpket=ppket_newrev}) are equivalent to 
\begin{eqnarray}
&&\hat{q}_{new}^\dag  | q \rangle_{new} =q | q \rangle_{new} , \label{qhatqket=qqket_new} \\
&&\hat{p}_{new}^\dag  | p \rangle_{new} =p | p \rangle_{new} , \label{phatpket=ppket_new} 
\end{eqnarray}
respectively.

A coherent state parametrized with a complex parameter $\lambda$ 
is defined up to a normalization factor by 
\begin{eqnarray}
| \lambda \rangle_{coh} &\equiv&  e^{\lambda a^\dag} | 0 \rangle \nonumber \\
&=& \sum_{n=0}^{\infty} \frac{\lambda^n}{\sqrt{n!}} | n \rangle , \label{lambda_coh_def}
\end{eqnarray}
and satisfies the relation 
\begin{equation}
a | \lambda \rangle_{coh} = \lambda | \lambda \rangle_{coh} . \label{a_lambda_ket=lambda_lambda_ket}
\end{equation}
In Eqs.(\ref{lambda_coh_def}) and (\ref{a_lambda_ket=lambda_lambda_ket}) $a$ and $a^\dag$ are annihilation and creation operators, which are 
defined in terms of $\hat{q}$ and $\hat{p}$ by 
\begin{eqnarray}
&&a = \sqrt{ \frac{m\omega}{2\hbar}}  \left( \hat{q} + i \frac{ \hat{p}}{m \omega}  \right) , \label{annihilation} \\
&&a^\dag = \sqrt{ \frac{m\omega}{2\hbar}}\left( \hat{q} - i \frac{ \hat{p}}{m \omega}  \right) \label{creation}.
\end{eqnarray}
The eigenstates of $\hat{q}$ and $\hat{p}$ are $| q \rangle$ and $| p \rangle$ for 
real $q$ and $p$ respectively, 
and they obey the following relations, 
\begin{eqnarray}
&&\hat{q} | q \rangle = q| q \rangle , \label{qhatqket=qqket} \\
&&\hat{p} | p \rangle = p | p \rangle ,    \label{phatpket=ppket} \\
&&\hat{q} | p \rangle = \frac{\hbar}{i} \frac{\partial}{\partial p}  | p \rangle , \label{qhatpket=ihbardeldelppket} \\ 
&&\hat{p} | q \rangle = i \hbar \frac{\partial}{\partial q}  | q \rangle ,  \label{phatqket=ihbardeldelqqket} \\
&&\langle q | p \rangle = \frac{1}{\sqrt{2\pi \hbar}} e^{\frac{i}{\hbar} pq} , \label{qbrapket=eihbarpq_real} \\ 
&&[ \hat{q},  \hat{p} ] = i \hbar. \label{commu_rel_qhatphat}
\end{eqnarray}
Substituting Eq.(\ref{annihilation}) for Eq.(\ref{a_lambda_ket=lambda_lambda_ket}), we 
obtain 
\begin{equation}
\left( \hat{q} + i \frac{\hat{p}}{m\omega} \right) | \lambda \rangle_{coh} 
= \sqrt{\frac{2\hbar}{m\omega}} \lambda | \lambda \rangle_{coh} .
\label{coh_q}
\end{equation}
Next we consider another coherent state $| \lambda' \rangle_{coh'}$ parametrized with a complex parameter $\lambda'$ 
of another harmonic oscillator defined with $m'$ and $\omega'$. 
Since the annihilation operator of the coherent state is given 
by replacing $m$ and $\omega$ with $m'$ and $\omega'$ in Eq.(\ref{annihilation}), 
\begin{equation}
a = \sqrt{ \frac{m' \omega'}{2\hbar}}  \left( \hat{q} + i \frac{ \hat{p}}{m' \omega'} \right) ,  
\label{annihilation'} 
\end{equation}
we obtain 
\begin{equation}
\left( \hat{p} + \frac{m' \omega'}{i} \hat{q} \right) | \lambda' \rangle_{coh'} = 
\frac{\lambda'}{i} \sqrt{2\hbar m' \omega'} | \lambda' \rangle_{coh'} .
\label{coh_p}
\end{equation}

Equations (\ref{coh_q}) and (\ref{coh_p}) inspire us to define 
$\hat{q}_{new}$, $\hat{p}_{new}$, $| q \rangle_{new}$ and $| p \rangle_{new}$ by 
\begin{eqnarray}
&&\hat{q}_{new} \equiv 
\frac{1}{ \sqrt{1 - \frac{m' \omega'}{m\omega} }  } \left( \hat{q} - i \frac{\hat{p}}{m\omega} \right), \label{def_qhat_new} \\
&&\hat{p}_{new} \equiv 
\frac{1}{ \sqrt{1 - \frac{m' \omega'}{m\omega} }  }  \left( \hat{p} - \frac{m' \omega'}{i} \hat{q} \right) , \label{def_phat_new} \\
&&| q \rangle_{new} 
\equiv 
\left\{ \frac{m\omega}{4\pi \hbar} \left( 1 - \frac{m'\omega'}{m\omega} \right) \right\}^\frac{1}{4} 
e^{- \frac{m\omega}{4\hbar} \left( 1 - \frac{m'\omega'}{m\omega} \right) {q}^2 }
| \sqrt{ \frac{m\omega}{2\hbar} 
\left( 1 - \frac{m'\omega'}{m\omega} \right) } q \rangle_{coh} ,  \nonumber \\
\\
&&| p \rangle_{new} 
\equiv 
\left( \frac{1 - \frac{m'\omega'}{m\omega} }{4\pi \hbar m' \omega' } \right)^{\frac{1}{4}} 
e^{ -\frac{1}{4 \hbar m' \omega'}  \left( 1 - \frac{m'\omega'}{m\omega} \right)  p^2 }
| i \sqrt{ \frac{1}{2\hbar m' \omega'} 
\left( 1 - \frac{m'\omega'}{m\omega} \right)}  p \rangle_{coh'} , \nonumber \\
\end{eqnarray}
with the introduction of complex $q$ and $p$ as 
\begin{eqnarray}
&&q \equiv \frac{1}{ \sqrt{1 - \frac{m' \omega'}{m\omega} } } 
\sqrt{\frac{2\hbar}{m\omega}} \lambda , 
\label{def_complexq_tildeq} \\
&&p \equiv 
\frac{\lambda'}{i} \sqrt{\frac{2\hbar m' \omega'}{1 - \frac{m'\omega'}{m\omega}}}. 
\label{def_complexp_tildep}
\end{eqnarray}
These expressions satisfy 
Eqs.(\ref{commutator_q_p_new})--(\ref{phatpket=ppket_new}), 
and $| q \rangle_{new}$ and $| p \rangle_{new}$ are normalized so that they satisfy 
the following relations, 
\begin{eqnarray}
{}_m\langle_{new}~ q' | q \rangle_{new} 
&=& \delta_c^{\epsilon_1} ( q' - q ) , \label{m_q'branew_qketnew}\\ 
{}_m\langle_{new}~ p' | p \rangle_{new} 
&=& \delta_c^{\epsilon'_1} ( p' - p ) , \label{m_p'branew_pketnew}
\end{eqnarray}
where we have used the expression of Eq.(\ref{delta_c_epsilon(q)}), and 
$\epsilon_1$ and $\epsilon'_1$ are given by 
\begin{eqnarray}
&&\epsilon_1 = \frac{\hbar}{m \omega \left( 1 - \frac{m'\omega'}{m\omega} \right)} ,
\label{epsilon_1} \\
&&\epsilon'_1 = \frac{\hbar m'\omega'}{1 - \frac{m'\omega'}{m\omega} } . \label{epsilon'_1}
\end{eqnarray}
For sufficiently large $m\omega$ 
the tamed delta function in Eq.(\ref{m_q'branew_qketnew}) converges 
for complex $q$ and $q'$ satisfying the condition like Eq.(\ref{cond_of_q_for_delta}),  
\begin{equation}
\left( \text{Re}(q-q') \right)^2 > \left( \text{Im}(q-q') \right)^2 . 
\label{cond_of_q'-q_for_delta}  
\end{equation}
This condition is satisfied only when $q$ and $q'$ are on the same path. 
Similarly, for sufficiently small $m' \omega'$ 
the tamed delta function in Eq.(\ref{m_p'branew_pketnew}) converges 
for complex $p$ and $p'$ satisfying the following condition, 
\begin{equation}
\left( \text{Re}(p-p') \right)^2 > \left( \text{Im}(p-p') \right)^2 . 
\label{cond_of_p'-p_for_delta} 
\end{equation}
Thus for sufficiently large $m\omega$ and small $m'\omega'$ 
Eqs.(\ref{m_q'branew_qketnew}) and (\ref{m_p'branew_pketnew}) represent 
the orthogonality relations for $| q \rangle_{new}$ and $| p \rangle_{new}$.

In the following we take $m\omega$ sufficiently large and $m'\omega'$ sufficiently small. 
Then we have the following relations for complex $q$ and $p$, 
\begin{eqnarray}
&&\int_C dq | q \rangle_{new} ~{}_m \langle_{new} q |  = 1 , \label{completion_complexq_ket2} \\
&&\int_C dp | p \rangle_{new} ~{}_m \langle_{new} p |  = 1 , \label{completion_complexp_ket2} \\
&&\hat{p}_{new}^\dag | q \rangle_{new} 
=i \hbar \frac{\partial}{\partial q} | q \rangle_{new}, \label{phatnewqketnew2} \\
&&\hat{q}_{new}^\dag | p \rangle_{new} 
= \frac{\hbar}{i} \frac{\partial}{\partial p} | p \rangle_{new}, \label{qhatnewpketnew2} \\
&&{}_m\langle_{new}~ q | p \rangle_{new} 
= \frac{1}{\sqrt{2 \pi \hbar}} \exp\left[\frac{i}{\hbar}p q \right], \label{basis_fourier_transf2}  \\
&&\langle p' | q \rangle_{new} 
=\frac{1}{ \sqrt{ 2\pi \hbar} } \exp\left[  - \frac{i}{\hbar} p' q \right] 
\quad \text{for real $p'$} , \label{brap'qketnew} \\
&&\langle q' | p \rangle_{new} 
=\frac{1}{ \sqrt{ 2\pi \hbar} } \exp\left[  \frac{i}{\hbar} q' p \right] 
\quad \text{for real $q'$} , \label{braq'pketnew} 
\end{eqnarray}
where Eqs.(\ref{completion_complexq_ket2}) and (\ref{completion_complexp_ket2}) represent 
the completeness relations for $|q \rangle_{new}$ and $|p \rangle_{new}$. 
Thus $\hat{q}_{new}$, $\hat{p}_{new}$, $| q \rangle_{new}$ and $| p \rangle_{new}$ 
with complex $q$ and $p$ 
obey the same relations as $\hat{q}$, $\hat{p}$, $| q \rangle$ and $| p \rangle$ with 
real $q$ and $p$ satisfy.

For real $q'$ and $p'$ $| q' \rangle_{new}$ and $| p' \rangle_{new}$ become 
$| q' \rangle$ and $| p' \rangle$ respectively, 
\begin{eqnarray}
&&| q' \rangle_{new} = | q' \rangle  , \label{qketnew_exp_qprimeketreal} \\
&&| p' \rangle_{new} = | p' \rangle . \label{pketnew_exp_pprimeketreal}
\end{eqnarray}
Also, $\hat{q}_{new}^\dag$ and $\hat{p}_{new}^\dag$ behave like hermitian operators 
$\hat{q}$ and $\hat{p}$ respectively for $| q' \rangle$ and $| p' \rangle$ with real $q'$ and $p'$,   
\begin{eqnarray}
&&\hat{q}_{new}^\dag | q' \rangle  =  \hat{q}  | q' \rangle  , \\
&&\hat{q}_{new}^\dag | p' \rangle  = \hat{q} | p' \rangle   , \\
&&
\hat{p}_{new}^\dag | q' \rangle =  \hat{p}  | q' \rangle  , \\
&&\hat{p}_{new}^\dag | p' \rangle   =  \hat{p}  | p' \rangle   . 
\end{eqnarray}
Using $| q \rangle_{new}$  we define a wave function $\psi(q)$ with complex $q$ by 
\begin{equation}
\psi(q) = {}_m\langle_{new}~ q | \psi \rangle . 
\end{equation}
When $q$ is real, this wave function becomes a usual one 
\begin{equation}
\psi(q) = \langle q | \psi \rangle . 
\end{equation}

One might feel a bit uneasy about the above replacement of the a priori correct operators 
$\hat{q}$ and $\hat{p}$ by 
$\hat{q}_{new}$ and $\hat{p}_{new}$. 
It might help a tiny bit to accept $\hat{q}_{new}$ and $\hat{p}_{new}$ 
to have in mind that 
operators smooth in $\hat{q}$ and $\hat{p}$ 
like $\hat{q}_{new}$ and $\hat{p}_{new}$ have generically eigenvalues 
filling the whole complex plane, while hermitian operators like $\hat{q}$ and $\hat{p}$ 
have eigenvalues only along a certain curve say on the real axis in the complex plane. 
The consolation for our replacement is then that 
for our purpose of having general contours running through eigenvalues 
we replaced the special operators $\hat{q}$ and $\hat{p}$ 
by the more generic ones $\hat{q}_{new}$ and $\hat{p}_{new}$. 
The philosophy should be that almost any small disturbance would anyway bring $\hat{q}$ and $\hat{p}$ 
into operators of the generic type with the whole complex plane as spectrum. 
The operators $\hat{q}_{new}$ and $\hat{p}_{new}$ are just concrete examples of such tiny deformation.
So we stress that the hermitian operators as $\hat{q}$ and $\hat{p}$ are special by 
having their eigenvalue spectrum on a curve say on the real axis in the complex plane 
rather than distributed all over it. 

Had we clung to the believe in curve-spectra, it would have been embarassing for our formalism that 
under Heisenberg time development one could have feared that the curve spectra would be transformed 
into new curve-spectra which might not match at the free choice of contour from time to time in our scheme. 
Now, however, 
as already stressed, if we use  $\hat{q}_{new}$ and $\hat{p}_{new}$, 
we have already from the beginning gone over to operators having any complex numbers as eigenvalues. 
So arbitrary deformation of the contour would a priori have no problem. 
Thus we claim that the contours of integration can be chosen 
freely at each time $t$, so that there is no need for any natural choice, 
which only has to run from $-\infty$ to $\infty$. 
%


In this section we have briefly reviewed the complex coordinate formalism of Ref.~\cite{Nagao:2011za}. 
See Ref.\cite{Nagao:2011za} for the detail of the formalism. 
For our convenience we show the summary of the comparison of the RAT and the CAT 
in table \ref{tab1:comparison}. 
In the next section we will study how the momentum and Hamiltonian 
are defined in the CAT based on the complex coordinate formalism. 
\begin{table}
\caption[AAA]{Various quantities in the RAT and the CAT}
\label{tab1:comparison}
\begin{center}
\begin{tabular}{|p{4cm}|p{4cm}|p{6.5cm}|}
\hline
    & the RAT & the CAT \\
\hline
parameters  &    $q$, $p$ real,   & $q$, $p$ complex \\  
\hline
complex conjugate  &    $*$   &  $*_{ \{~\}}$  \\  
\hline
hermitian conjugate  &    $\dag$   &  $\dag_{ \{~\}}$  \\  
\hline
delta function of $q$ &  $\delta(q)$ defined for    & $\delta_c(q)$ defined for $q$ s.t.\\
& real $q$ &  $\left( \text{Re}(q) \right)^2 > \left( \text{Im}(q) \right)^2$  \\ 
\hline
bras of $| q\rangle$, $| p\rangle$   &  $\langle q | = ( | q\rangle )^\dag $,  & 
${}_m\langle_{new}~ q |=\langle_{new}~ q^* |  = ( | q \rangle_{new} )^{\dag_q}  $,  \\
& $\langle p | = ( | p\rangle )^\dag$ & ${}_m\langle_{new}~ p |=\langle_{new}~ p^* |  = ( | p \rangle_{new} )^{\dag_p}  $ \\

\hline  
completeness for  &  $\int_{-\infty}^\infty |q\rangle \langle q | dq =1$ , & 
$\int_C |q \rangle_{new} ~{}_m\langle_{new}~ q | dq =1$ , \\
$| q \rangle$ and $|p\rangle$ &  $\int_{-\infty}^\infty |p \rangle \langle p | dp =1$  & 
$\int_C |p\rangle_{new} ~{}_m\langle_{new}~ p | dp =1$ \\
& along real axis & $C$: any path running from $-\infty$ to $\infty$\\
\hline 
orthogonality for  &  $\langle q | q'\rangle = \delta(q-q')$ , & 
${}_m\langle_{new}~ q | q'\rangle_{new} = \delta_c (q-q')$ , \\
$| q\rangle$ and $|p\rangle$ &  $\langle p | p'\rangle = \delta(p-p')$ & 
${}_m\langle_{new}~ p | p'\rangle_{new} = \delta_c (p-p')$ \\
\hline 
basis of Fourier expansion &  $\langle q | p\rangle = \exp(ipq)$ & 
${}_m\langle_{new}~ q | p\rangle_{new} = \exp(ipq)$ \\
\hline 
$q$ representation of  $| \psi \rangle$  & $\psi(q) = \langle q | \psi \rangle$ & 
$\psi(q) = {}_m\langle_{new}~ q | \psi \rangle $ \\
\hline 
complex conjugate of  $\psi(q)$ & $\langle q | \psi \rangle^*=\langle \psi | q \rangle$ & 
${}_m\langle_{new}~ q | \psi \rangle^{*_q } =\langle \psi | q \rangle_{new}$ \\
\hline
normalization of $\psi(q)$  & $\int_{-\infty}^\infty \psi(q)^* \psi(q) dq = 1$ & 
$\int_C \psi(q)^{*_q } \psi(q) dq = 1$ \\
\hline
\end{tabular}
\end{center}
\end{table}
%
%

\section{The derivation of the Hamiltonian and the momentum relation}\label{derivation_hamiltonian}

In this section we study how the momentum and the Hamiltonian are defined in the CAT. 
Before starting the explicit analysis we mention one of the general 
aspects of the CAT.

\subsection{A condition on the mass}

As a simple case of the CAT 
we consider a non-relativistic particle with position $q$ in one dimension with a potential 
energy $V(q)$, which can now be a complex function of the real variable $q$ at first. 
Also the mass $m$ of this particle can be complex. 
The Lagrangian is given by  
\begin{equation}
L =\frac{1}{2}m \dot{q}^2- V(q) ,
\end{equation}
and the functional integral takes the form 
\begin{eqnarray}
&&\int_C e^{\frac{i}{\hbar} \int L(q, \dot{q}) dt } {\cal D} q  \nonumber\\    
&=&\int_C \exp\left[ \frac{i}{\hbar} \sum_t \left\{ \frac{1}{2} m \left( \frac{q_{t+dt} - q_t}{\Delta t} \right)^2 
- V(q_t) \right\} \Delta t \right] \Pi_t \left( dq_t \frac{m}{2\pi i \hbar \Delta t} \right) , \nonumber \\
\end{eqnarray}
where discretizing the time direction we have written $\dot{q}$ as 
\begin{equation}
\dot{q} = \frac{ q(t+ dt) - q(t) }{dt} , \label{dot_q} 
\end{equation}
and introduced the notation $q_t$ and $q_{t+dt}$ for $q(t)$ and $q(t+dt)$ as 
\begin{eqnarray}
&&q_t \equiv q(t) , \\
&&q_{t+dt} \equiv q(t+dt) .
\end{eqnarray}
We regard $q_t$ and $q_{t+dt}$ as independent variables.
We consider a theory such that the asymptotic values of dynamical variables such as $q$ or $p$ are on the real axis.  
The path $C$ denotes a set of arbitrary paths running from $-\infty$ to $\infty$ in the complex plane 
at each time $t$. 
Since this regularized expression is analytical in all the $q_t$ for all different moments $t$, 
we can arbitrarily deform the contour for each $q_t$ independent of each other, 
except for keeping some continuity in $t$.

In the Lagrangian, since $dt$ is taken to be a small quantity 
in the denominator of $\dot{q}$, 
the kinetic term $\frac{1}{2} m \dot{q}^2$ tends to 
dominate over the potential term $V(q)$. 
Therefore a kind of restriction is needed for 
the kinetic term. 
In practice we impose on the mass the condition 
\begin{equation}
\text{Re}\left( \frac{i}{\hbar} \frac{m}{2}  \right)  < 0 \quad \Leftrightarrow \quad m_I>0 ,
\end{equation}
where $m_I$ is the imaginary part of $m$. 
Then we can prevent the kinetic term in the integrand 
from blowing up for $\dot{q}\rightarrow \pm\infty$ along real axis. 
We note that $m_I=0$ can be accommodated with this condition, since 
$m_I=0$ is obtained by taking an infinitesimally small limit of positive $m_I$. 
Therefore in the following we impose on $m$ the condition 
\begin{equation}
m_I \geq 0 . \label{m_I_positive_cond}
\end{equation}
This is the wanted condition for $m_I$ in order that FPI of the CAT is meaningful.

\subsection{Our approach}

We first consider the momentum. 
In classical mechanics of the RAT the momentum is defined by the relation
\begin{equation}
p=\frac{\partial L}{\partial \dot{q}} . \label{def_p_delLdelqdot} 
\end{equation}
Can we use Eq.(\ref{def_p_delLdelqdot}) as the definition of the momentum also in the CAT? 
It is not clear whether we can or not from a point of view
of quantum mechanics because it includes $q_{t+dt}$, a quantity at time $t+dt$. 
The quantity $q_{t+dt}$ is somehow unclear unless 
we define it properly including the fluctuation in the time-development from 
a quantity at time $t$. 
Hence we attempt to describe $q_{t+dt}$ via FPI 
and derive the relation (\ref{def_p_delLdelqdot}) 
from FPI in the CAT.
In FPI the time-development of some wave function 
${}_m \langle_{new}~ q_t | \psi(t) \rangle$ at time $t$ to time $t+dt$ is described by  
\begin{equation}
{}_m \langle_{new}~ q_{t+dt} | \psi(t+dt) \rangle = \frac{1}{\alpha} \int_C e^{\frac{i}{\hbar} \Delta t L(q,\dot{q})}
{}_m \langle_{new}~ q_{t} | \psi(t) \rangle d q_t , \label{time_dev_qbraAket}
\end{equation}
where $\frac{1}{\alpha}$ is a $dt$-dependent normalization factor. 
For the purpose to derive the relation (\ref{def_p_delLdelqdot}) 
we take the following approach. 
First we imagine that we have some wave function 
${}_m \langle_{new}~ q_t | \xi \rangle$ with some number $\xi$ which obeys 
\begin{eqnarray}
{}_m \langle_{new}~ q_t | \hat{p}_{new} | \xi \rangle &=&
\frac{\hbar}{i}\frac{\partial}{\partial q_t} ~{}_m \langle_{new}~ q_t | \xi \rangle \nonumber \\
&=&
\frac{\partial L}{\partial \dot{q}}\left( q_t, \frac{\xi-q_t}{dt} \right) 
{}_m \langle_{new}~ q_t | \xi \rangle . \label{phat_p_xi2}
\end{eqnarray}
For fixed $\xi$ Eq.(\ref{phat_p_xi2}) is a differential equation to be obeyed by 
${}_m \langle_{new}~ q_t | \xi \rangle$. 
This is an eigenvalue problem of the momentum operator. 
If the set $\left\{ |\xi \rangle \right\}$ could be an approximately reasonable basis which has 
at least roughly completeness and orthogonality, 
\begin{eqnarray}
&&1 \simeq \int_C d\xi | \xi \rangle ~{}_m \langle \xi | , \\
&& {}_m \langle \xi | \xi' \rangle \simeq \delta_c(\xi -\xi') ,
\end{eqnarray}
we could expand the wave function ${}_m \langle_{new}~ q_t | \psi(t) \rangle$ into 
a linear combination of ${}_m \langle_{new}~ q_t | \xi \rangle$ as 
\begin{eqnarray}
{}_m \langle_{new}~ q_t | \psi(t) \rangle 
&\simeq& \int_C d\xi  ~{}_m \langle_{new}~ q_t | \xi \rangle ~{}_m \langle \xi | \psi(t) \rangle \nonumber \\
&=& \int_C d\xi ~{}_m \langle_{new}~ q_t | \psi(t) \rangle|_\xi, \label{qbra_Atket_exp_lc_xi}
\end{eqnarray}
where in the second equality for our convenience we have introduced 
the notation ${}_m \langle_{new}~ q_t | \psi(t) \rangle|_\xi$, 
which is the contributed part of ${}_m \langle_{new}~ q_t | \psi(t) \rangle$ from $\xi$-component. 
By solving Eq.(\ref{phat_p_xi2}) and using the expansion (\ref{qbra_Atket_exp_lc_xi}) 
we attempt to show that Eq.(\ref{time_dev_qbraAket}) would give 
\begin{equation}
{}_m \langle_{new}~ q_{t+dt} | \psi(t+dt) \rangle|_\xi \propto \delta_c (q_{t+dt}-\xi ), \label{prompt_delta}
\end{equation}
because this relation means that only the component with $\xi \simeq q_{t+dt}$ 
contributes a non-zero value to ${}_m \langle_{new}~ q_{t+dt} | \psi(t+dt) \rangle$. 
The relation (\ref{def_p_delLdelqdot}) can be obtained by showing Eq.(\ref{prompt_delta}). 
In the following we take this strategy, and calculate 
${}_m \langle_{new}~ q_{t+dt} | \psi(t+dt) \rangle$ explicitly. 
By expressing this quantity in terms of $| \psi(t) \rangle$ and looking at its time-development operator, 
we can obtain the Hamiltonian and Schr\"{o}dinger equation.

\subsection{A general prescription of our approach}
%

Before going ahead we mention that the above approach can be presented in a more general prescription. 
Not only the momentum but also other operators including the momentum can be dealt with 
as follows. 
We consider relations involving a wave function and a parameter $\xi$ going into the Lagrangian 
at place of $q_{t+dt}$ as Eq.(\ref{phat_p_xi2}). 
We are thus motivated to think about a family of $\xi$-parameterized operators $A(\xi)$ 
which operates on a wave function ${}_m \langle_{new}~ q_t | \xi \rangle$ to give a vanishing value as 
\begin{equation} 
A(\xi) ~{}_m \langle_{new}~ q_t | \xi \rangle = 0 . \label{A_xi_ket}
\end{equation}
If $|\xi \rangle $ satisfies the following relation 
\begin{equation}
\int_C dq_t e^{ \frac{i}{\hbar} L dt }    {}_m \langle_{new}~ q_{t} | \xi \rangle \propto \delta_c(\xi- q_{t+dt} ) ,
\end{equation}
$A(\xi)$ acting on ${}_m \langle_{new}~ q_t | \xi \rangle$ 
effectively projects out the $\xi=q_{t+dt}$ component. 
The strategy explained in the previous subsection is an example of this general prescription.  
Indeed if we take 
$A(\xi) = \frac{\hbar}{i}\frac{\partial}{\partial q_t} - \frac{\partial L}{\partial \dot{q}}\left( q_t, \frac{\xi - q_t}{dt} \right) $, 
which is the difference between the two operators acting on both sides in Eq.(\ref{phat_p_xi2}), 
Eq.(\ref{phat_p_xi2}) is reproduced from Eq.(\ref{A_xi_ket}). 
For other purpose of studying the Hamiltonian, for example, 
$A(\xi)$ could be taken as 
$A(\xi) =-\frac{\hbar^2}{2m} \frac{\partial^2 }{\partial^2 q_t}
- \frac{1}{2m}  \left( \frac{\partial L}{\partial \dot{q}}\left( q_t, \frac{\xi - q_t}{dt}  \right) \right)^2$.

\subsection{The wave function solution to the eigenvalue problem}

Based on the approach explained in the foregoing subsections we shall solve the 
eigenvalue problem of Eq.(\ref{phat_p_xi2}). 
Usually the Lagrangian $L$ is written as $L(q,\dot{q})=T-V$, 
where $T$ is a kinetic term and $V$ is a potential term. 
Taylor-expansion of $L(q,\dot{q})$ with regard to $\dot{q}$ is written as 
\begin{eqnarray}
L(q,\dot{q}) 
&=& L|_{\dot{q}=0} + 
\dot{q}\frac{\partial L}{\partial \dot{q}}\bigg|_{\dot{q}=0} 
+\frac{1}{2}  \dot{q}^2 \frac{\partial^2 L}{\partial \dot{q}^2 }\bigg|_{\dot{q}=0} \nonumber \\
&=&
\frac{1}{2}  \frac{\partial^2 L}{\partial \dot{q}^2 } \bigg|_{\dot{q}=0} \frac{1}{(dt)^2} q_t^2 
-\left( \frac{\partial^2 L}{\partial \dot{q}^2 }\bigg|_{\dot{q}=0}  \frac{q_{t+dt}}{(dt)^2} 
+\frac{\partial L}{\partial \dot{q} }\bigg|_{\dot{q}=0}   \frac{1}{dt} \right) q_t 
+ \frac{1}{2}\frac{\partial^2 L}{\partial \dot{q}^2 }\bigg|_{\dot{q}=0}  \frac{q_{t+dt}^2}{(dt)^2} 
\nonumber \\
&&
+ \frac{\partial L}{\partial \dot{q} }\bigg|_{\dot{q}=0}  \frac{q_{t+dt}}{dt}
+ L|_{\dot{q}=0}  , \label{taylor_L} 
\end{eqnarray}
where in the first equality we have assumed the normal Lagrangian such that 
$\frac{\partial^3 L}{\partial \dot{q}^3}=0$ and in the second equality 
we have used Eq.(\ref{dot_q}). 
We consider the case where $T$ is given by 
\begin{equation}
T=\frac{m}{2} \dot{q}^2 .
\end{equation}
Then $\frac{\partial L}{\partial \dot{q}} \left( q,\dot{q} \right)$ is written as 
\begin{equation}
\frac{\partial L}{\partial \dot{q}} \left( q, \dot{q} \right) 
= m \dot{q} =\frac{m( q_{t+dt}-q_{t} )}{\Delta t} , 
\end{equation}
so we have 
\begin{equation}
\frac{\partial L}{\partial \dot{q}}\left( q_t, \frac{\xi -q_t}{dt} \right) 
= \frac{m(\xi-q_t)}{\Delta t} . \label{par_L_par_dotq_xi}
\end{equation} 
Substituting Eq.(\ref{par_L_par_dotq_xi}) for Eq.(\ref{phat_p_xi2}), we find that 
the solution of Eq.(\ref{phat_p_xi2}) is given by 
\begin{eqnarray}
{}_m\langle_{new}~ q_t | \xi \rangle &=& C_A \exp\left[ -C_1 (q_t - \xi)^2 \right]  \nonumber \\ 
&=& C_A \tilde{\psi}_\xi (q_t) , \label{qbra_Atket_xi}
\end{eqnarray}
where 
\begin{equation}
C_1=\frac{im}{2\hbar \Delta t}
\end{equation}
and $C_A$ is a normalization factor. 
Here we note that the above wave function is normalizable for $m_I  \leq 0$, 
but not for $m_I > 0$. 
Namely the wave function diverges except for $m_I=0$ case 
when the condition (\ref{m_I_positive_cond}), 
which is generally required for convergence of the CAT, is satisfied.
It means that the wave function destined to make $q_{t+dt}$ be $\xi$ by being 
a solution to Eq.(\ref{phat_p_xi2}) is not a good wave function, but rather a non-normalizable one. 
However, we think that it would not make a big problem because 
if we consider a test function which converges sufficiently, 
the product of the wave function and the test function in an integrand can converge. 
We will come back to this point later.

Next we explain how we normalize the wave function ${}_m\langle q_t | \xi \rangle $. 
Since we have the following relation, 
\begin{eqnarray}
&&\int_C \tilde{\psi}_{\xi'} (q_t)^{*_{q_t,\xi'}}  \tilde{\psi}_\xi (q_t) dq_t \nonumber \\
&=& 
\int_C \tilde{\psi}_{\xi'^*} (q_t^*)^*  \tilde{\psi}_\xi (q_t) dq_t \nonumber \\
&=& 
\int_C dq_t
\exp\left[ -\frac{i(m-m^*)}{2\hbar \Delta t} q_t^2 
+\frac{i(m\xi - m^* \xi')}{\hbar \Delta t} q_t
-\frac{i(m\xi^2 - m^* \xi'^2)}{2\hbar \Delta t} 
\right]  , \label{qbra_Atket_xi3}
\end{eqnarray}
we calculate the normalization factor according to whether $m$ is real or not.

\subsubsection{In the case of real $m$}

In the case of real $m$, namely $m^*=m$, we have 
\begin{eqnarray}
\int_C \tilde{\psi}_{\xi'} (q_t)^{*_{q_t,\xi'}}  \tilde{\psi}_\xi (q_t) dq_t 
&=&
\int_C \exp\left[ \frac{im(\xi - \xi')}{\hbar \Delta t} q_t
-\frac{i m (\xi^2 - \xi'^2)}{2\hbar \Delta t} 
\right]  dq_t \nonumber \\
&=&
\frac{ 2 \pi \hbar \Delta t }{m} \delta_c (\xi' - \xi)
\exp\left[-\frac{i m (\xi^2 - \xi'^2)}{2\hbar \Delta t} 
\right]  \nonumber \\
&=&
\frac{ 2 \pi \hbar \Delta t }{m} \delta_c (\xi' - \xi) , \label{qbra_Atket_xi6}
\end{eqnarray}
where $\delta_c(\xi' - \xi)$ is well-defined for $\xi$ and $\xi'$ such that 
$\left\{ \text{Re}(\xi' - \xi) \right\}^2 > \left\{ \text{Im}(\xi' - \xi) \right\}^2$. 
Therefore we can normalize the wave function by choosing $C_A$ as 
\begin{equation}
C_A= \sqrt{ \frac{ m}{ 2 \pi \hbar \Delta t} } .
\end{equation}
Thus we have 
\begin{eqnarray} 
\int_C ( {}_m  \langle_{new}~ q_t | \xi' \rangle)^{*_{q_t,\xi'}} {}_m \langle_{new}~ q_t | \xi \rangle dq_t
&=& 
\int_C {}_m \langle \xi'  | q_t \rangle_{new}~ ~{}_m \langle_{new}~ q_t | \xi \rangle dq_t \nonumber \\
&=& \delta_c (\xi' -\xi) , \label{qbra_Atket_xi8}
\end{eqnarray}
and we can show  
\begin{eqnarray}
\int_C ({}_m \langle_{new}~ q'_t | \xi \rangle)^{*_{q_t,\xi'}} {}_m \langle_{new}~ q_t | \xi \rangle d\xi 
&=& \int_C {}_m \langle \xi  | q'_t \rangle_{new}~ ~{}_m \langle_{new}~ q_t | \xi \rangle d\xi \nonumber \\
&=& \delta_c (q_t' -q_t) . \label{qbra_Atket_xi8a}
\end{eqnarray}
From Eqs.(\ref{qbra_Atket_xi8}) and (\ref{qbra_Atket_xi8a}) we obtain 
\begin{eqnarray}
&&{}_m \langle \xi' | \xi \rangle =\delta_c (\xi' -\xi) , \\
&&\int_C | \xi \rangle ~{}_m \langle \xi  | d\xi =1 . 
\end{eqnarray}
Therefore we see that in the case of real $m$ $| \xi \rangle$ has both of the completeness 
and orthogonority, and thus 
${}_m \langle_{new}~ q_t | \xi \rangle$ can work as a basis.

\subsubsection{In the case of complex $m$}

In the case of $m^*\neq m$, we have 
\begin{eqnarray}
&&\int_C \tilde{\psi}_{\xi'} (q_t)^{*_{q_t,\xi'}}  \tilde{\psi}_\xi (q_t) dq_t \nonumber \\
&=&
\int_C dq_t 
\exp
\left[ -\frac{i(m-m^*)}{2\hbar \Delta t} 
\left\{ q_t - \frac{m\xi - m^* \xi' }{m - m^*} \right\}^2 
+ \frac{i |m|^2 (\xi - \xi')^2 }{2\hbar \Delta t (m - m^*) } \right]. \nonumber \\
\label{qbra_Atket_xi4}
\end{eqnarray}
We note that the integral is divergent for $m_I>0$, so we cannot perform the Gaussian 
integral and normalize the wave function for $m_I>0$. 
To improve this situation we propose to introduce 
another special state 
$| \text{anti} ~ \xi \rangle$ 
such that we can have both of the completeness and orthogonality. 
First we try to find a dual basis, a set ${}_m\langle \text{anti} ~ \xi |$ such that 
\begin{equation}
\int_C | \xi \rangle ~{}_m\langle \text{anti} ~ \xi | d\xi = 1 ,  
\end{equation}
which is rewritten as 
\begin{equation}
\int_C {}_m\langle_{new}~ q | \xi \rangle ~{}_m\langle \text{anti} ~ \xi | q' \rangle_{new}~ d\xi 
= \delta_c(q-q') . \label{anti-bara-xi}
\end{equation}  
From the expression of ${}_m\langle_{new}~ q_t | \xi \rangle$ in Eq.(\ref{qbra_Atket_xi}), 
we consider the following form of ${}_m\langle \text{anti} ~ \xi | q_t' \rangle_{new}$, 
\begin{equation}
{}_m\langle \text{anti} ~ \xi | q_t' \rangle_{new}~ = C_B^{*} \exp\left[ C_1(q_t' -\xi)^2 \right] , 
\label{bra_antixi_q_ket}
\end{equation}
from which we have 
\begin{eqnarray} 
{}_m\langle_{new}~ q_t' | \text{anti} ~\xi \rangle &=& 
( {}_m\langle \text{anti} ~ \xi | q_t' \rangle_{new} )^{*_{q_t',\xi}} \nonumber \\
&=& C_B \exp\left[ C_1^{*} (q_t' - \xi)^2 \right]  .  \label{bra_antixi_q_ket2}
\end{eqnarray}
Since $C_1^{*}$ is written as 
\begin{equation}
C_1^{*}=-\frac{im^*}{2\hbar \Delta t} , 
\end{equation}
${}_m\langle_{new}~ q_t' | \text{anti} ~\xi \rangle$ is normalizable for $m_I > 0$. 
Substituting Eqs.(\ref{qbra_Atket_xi}) and (\ref{bra_antixi_q_ket}) for 
Eq.(\ref{anti-bara-xi}), we obtain 
\begin{equation}
C_B^* C_A = \frac{m}{2\pi \hbar \Delta t } . \label{C_B*C_A}
\end{equation}
Then we can also show 
\begin{equation}
\int_C {}_m\langle_{new}~ q_t | \xi' \rangle ~{}_m\langle \text{anti} ~ \xi | q_t \rangle_{new}~ dq_t = 
\delta_c (\xi-\xi') . 
\label{anti-bara-xi2}
\end{equation}  

In Eq.(\ref{C_B*C_A}) we have ambiguity or freedom to distribute normalization factor 
between ${}_m\langle \text{anti} ~ \xi |$ and $| \xi \rangle$, 
so no unique normalization is required.  
However, if we choose $C_A$ and $C_B$ as 
\begin{eqnarray}
&&C_A = \sqrt{ \frac{m}{2\pi \hbar \Delta t } } , \label{C_Acomplexm} \\ 
&&C_B = \sqrt{ \frac{m^*}{2\pi \hbar \Delta t } } , \label{C_Bcomplexm} 
\end{eqnarray}
then the wave functions of Eqs.(\ref{qbra_Atket_xi}) and (\ref{bra_antixi_q_ket2}) 
are explicitly written as 
\begin{eqnarray}
&&{}_m\langle_{new}~ q_t | \xi \rangle =\sqrt{ \frac{m}{2\pi \hbar \Delta t } } 
 \exp\left[ -C_1 (q_t - \xi)^2 \right] , \\
&& {}_m\langle_{new}~ q_t | \text{anti} ~\xi \rangle =\sqrt{ \frac{m^*}{2\pi \hbar \Delta t } } 
 \exp\left[ C_1^{*} (q_t - \xi)^2 \right] , 
\end{eqnarray}
and we note that with the choice of Eqs.(\ref{C_Acomplexm}) and (\ref{C_Bcomplexm}) 
${}_m\langle \text{anti} ~\xi |$ corresponds to 
${}_m\langle \xi |$ when we take $m$ real. 
Therefore the normalization of Eqs.(\ref{C_Acomplexm}) and (\ref{C_Bcomplexm}) 
seems to be a natural choice.

From the above consideration we can say that in the case of complex $m$ 
we have both of the orthogonality and completeness relation 
for $|\xi \rangle$ by introducing ${}_m\langle \text{anti} ~ \xi |$, 
and thus ${}_m \langle_{new}~ q_t | \xi \rangle$ can roughly work as a basis. 
Due to the introduction of ${}_m\langle \text{anti} ~\xi |$, 
Eq.(\ref{qbra_Atket_exp_lc_xi}) is replaced with 
\begin{eqnarray}
{}_m \langle_{new}~ q_t | \psi(t) \rangle 
&\simeq& \int_C d\xi  ~{}_m \langle_{new}~ q_t | \xi \rangle 
~{}_m \langle \text{anti} ~ \xi | \psi(t) \rangle \nonumber \\
&=& \int_C d\xi ~{}_m \langle_{new}~ q_t | \psi(t) \rangle|_\xi. \label{qbra_Atket_exp_lc_xi2}
\end{eqnarray}
This replacement is applied not only in the complex $m$ case but also in the real $m$ case 
since ${}_m \langle \text{anti} ~ \xi |$ corresponds to ${}_m \langle \xi |$ in the 
latter case as mentioned above.

\subsection{Explicit estimation}

In the previous subsection we have obtained the explicit form of 
${}_m \langle_{new}~ q_t | \xi \rangle$, which is the solution to 
the eigenvalue problem of Eq.(\ref{phat_p_xi2}), and showed that 
$| \xi \rangle$ has both of the orthogonality and completeness relation. 
We are thus prepared to calculate 
${}_m \langle_{new}~ q_{t+dt} | \psi(t+dt) \rangle =\frac{1}{\alpha} \int_C e^{\frac{i}{\hbar} \Delta t L(q_t,\dot{q_t})}
{}_m \langle_{new}~ q_t | \psi(t) \rangle dq_t$ explicitly. 
Substituting Eqs.(\ref{taylor_L}), (\ref{qbra_Atket_xi}) and (\ref{qbra_Atket_exp_lc_xi2}) 
for Eq.(\ref{time_dev_qbraAket}), we have
\begin{eqnarray}
&&{}_m \langle_{new}~ q_{t+dt} | \psi(t+dt) \rangle|_\xi \nonumber \\ 
&=& \frac{1}{\alpha} \int_C e^{\frac{i}{\hbar} \Delta t L(q_t,\dot{q_t})}
{}_m \langle_{new}~ q_t | \psi(t) \rangle|_\xi dq_t \nonumber \\
&=&
\frac{C_A}{\alpha}  ~{}_m \langle \text{anti} ~\xi | \psi(t) \rangle 
\int_C \exp\left[  \frac{i}{2\hbar dt} 
\left( \frac{\partial^2 L}{\partial \dot{q}^2 } \bigg|_{\dot{q}=0} -m \right) q_t^2 
- \frac{i}{\hbar} 
\left( 
\frac{1}{dt} \frac{\partial^2 L}{\partial \dot{q}^2 } \bigg|_{\dot{q}=0} q_{t+dt}
\right. \right.
\nonumber \\
&&
\left. \left. 
+\frac{\partial L}{\partial \dot{q} } \bigg|_{\dot{q}=0}  
- \frac{m\xi}{dt} \right) q_t 
+ \frac{i}{2\hbar} \frac{\partial^2 L}{\partial \dot{q}^2 }\bigg|_{\dot{q}=0} 
\frac{q_{t+dt}^2}{dt} 
+ \frac{i}{\hbar} \frac{\partial L}{\partial \dot{q} }\bigg|_{\dot{q}=0} q_{t+dt}
+\frac{i}{\hbar}  L|_{\dot{q}=0}  dt 
-\frac{im}{2\hbar \Delta t} \xi^2
\right]
dq_t . \nonumber \\
\label{time_dev_qbraAket2}
\end{eqnarray}
We shall evaluate Eq.(\ref{time_dev_qbraAket2}) first in the free case, 
and next in the potential case.

\subsubsection{Free case}

First we consider the free case, $V=0$. 
In this case, since each component of the Taylor-expanded $L$ in Eq.(\ref{taylor_L}) is given by 
\begin{eqnarray}
&&L(q=q,\dot{q}=0)  = -V= 0 , \\
&&\frac{\partial L}{\partial \dot{q}}\bigg|_{\dot{q}=0}  = m\dot{q}|_{\dot{q}=0}  =0 , \\
&&\frac{\partial^2 L}{\partial \dot{q}^2 }\bigg|_{\dot{q}=0} = m, 
\end{eqnarray}
Eq.(\ref{time_dev_qbraAket2}) is calculated as 
\begin{eqnarray}
&&{}_m \langle_{new}~ q_{t+dt} | \psi(t+dt) \rangle|_\xi \nonumber \\ 
&=&
\frac{C_A}{\alpha}  ~{}_m \langle \text{anti} ~\xi | \psi(t) \rangle 
\int_C \exp\left[  - \frac{i}{\hbar} 
\frac{m}{dt} ( q_{t+dt} -\xi ) q_t 
+ \frac{i m}{2\hbar} \frac{q_{t+dt}^2}{dt} 
-\frac{im}{2\hbar \Delta t} \xi^2
\right] dq_t \nonumber \\
&=&
\frac{C_A}{\alpha}  ~{}_m \langle \text{anti} ~\xi | \psi(t) \rangle 
2\pi\hbar \delta_c(\xi - q_{t+dt}) \frac{dt} {m}
\exp\left[  \frac{i m}{2\hbar dt}  (q_{t+dt}^2 - \xi^2 ) \right] .
\label{time_dev_qbraAket_k=0}
\end{eqnarray}
The above result means that 
\begin{equation}
{}_m \langle_{new}~ q_{t+dt} | \psi(t+dt) \rangle|_\xi \propto \delta_c (q_{t+dt} -\xi) .
\end{equation} 
Therefore unless $\xi=q_{t+dt}$ the value of ${}_m \langle_{new}~ q_{t+dt} | \psi(t+dt) \rangle|_\xi$ vanishes. 
I.e. only the component with $\xi=q_{t+dt}$ contributes to 
the value of ${}_m \langle_{new}~ q_{t+dt} | \psi(t+dt) \rangle$. 
Thus we have obtained the momentum relation of Eq.(\ref{def_p_delLdelqdot}) 
from the FPI point of view.


Next we evaluate ${}_m \langle_{new}~ q_{t+dt} | \psi(t+dt) \rangle$. 
This quantity is calculated as 
\begin{eqnarray}
&&{}_m \langle_{new}~ q_{t+dt} | \psi(t+dt) \rangle\nonumber \\ 
&=&
\int_C d\xi ~{}_m \langle_{new}~ q_{t+dt} | \psi(t+dt) \rangle|_\xi \nonumber \\
&=&
\frac{C_A}{\alpha} ~{}_m \langle \text{anti} ~\xi=q_{t+dt} | \psi(t) \rangle 
2\pi\hbar \frac{dt} {m} \nonumber \\
&=&
\frac{1}{\alpha} \frac{2\pi\hbar dt}{m} C_A   C_B^*
\int_C dq_t 
\exp\left[ \frac{ im}{2\hbar \Delta t} (q_t -q_{t+dt})^2 \right]
~{}_m \langle_{new}~ q_t | \psi(t) \rangle  \nonumber \\
&\simeq& 
\frac{1}{\alpha}\sqrt{ \frac{ 2\pi i \hbar \Delta t}{m} }
\left\{
{}_m \langle_{new}~ q_{t+dt} | \psi(t) \rangle
-\frac{i}{\hbar} 
\left( -\frac{\hbar^2}{2m} \right) \Delta t 
\frac{\partial^2}{\partial q_{t+dt}^2}  
~{}_m \langle_{new}~ q_{t+dt} | \psi(t) \rangle 
\right\} 
\nonumber \\
&\simeq& \frac{1}{\alpha}\sqrt{ \frac{ 2\pi i \hbar \Delta t}{m} }
{}_m \langle_{new}~ q_{t+dt} |  
\exp\left[ -\frac{i}{\hbar} \hat{H_0}  \Delta t \right] |\psi(t) \rangle , 
\label{time_dev_qbraAket_k=0_2}
\end{eqnarray}
where in the third equality we have inserted 
\begin{equation}
1=\int_C dq_t | q_t \rangle_{new} ~{}_m \langle_{new}~ q_t | , \label{qt_completeness}
\end{equation}
and used the following expression, 
\begin{eqnarray}
{}_m \langle \text{anti} ~\xi=q_{t+dt} | q_t \rangle_{new}~ &=& 
{}_m \langle_{new}~ q_t |  \text{anti} ~\xi=q_{t+dt} \rangle^{*_{q_t, q_{t+dt}}} \nonumber \\
&=& \left( C_B \exp\left[ -\frac{im^*}{2\hbar \Delta t} (q_t-q_{t+dt})^2 \right] \right)^{*_{q_t, q_{t+dt}}} 
\nonumber \\
&=& C_B^* \exp\left[ \frac{im}{2\hbar \Delta t} (q_t-q_{t+dt})^2 \right] . \label{comp_conj_exp}
\end{eqnarray}
In the fourth equality of Eq.(\ref{time_dev_qbraAket_k=0_2}), 
noting that the exponential converges for $m_I \geq 0$, 
we have used the following relation 
\begin{equation}
\sqrt{\frac{\alpha}{\pi}} \exp\left[ -\alpha(x-\beta)^2 \right] 
\simeq \delta_c(x-\beta) + \frac{1}{4\alpha} \frac{\partial^2}{\partial x^2} \delta_c(x-\beta) 
+ \cdots  , \label{exp_delta_func}
\end{equation}
which stands for large positive $\text{Re} \alpha$, and integrated them out with regard to $q_t$. 
In the last expression $\hat{H_0} $ is given by 
\begin{equation}
\hat{H_0} = \frac{1}{2m} \hat{p}_{new}^2 . \label{H_0}
\end{equation}
Here we take $\alpha= \sqrt{\frac{2\pi i \hbar dt}{m}}$ so that 
both sides  of Eq.(\ref{time_dev_qbraAket_k=0_2}) correspond to each other in the vanishing limit of $dt$.
Equations (\ref{time_dev_qbraAket_k=0_2}) and (\ref{H_0}) show that we have obtained 
the Hamiltonian $\hat{H_0}$ from the Lagrangian $L$ via the FPI point of view 
in the free case. 
In addition, Eq.(\ref{time_dev_qbraAket_k=0_2}) is reduced to 
$|\psi(t+dt) \rangle = e^{-\frac{i}{\hbar} \hat{H}_0 dt}|\psi(t) \rangle$. 
Thus we have derived the Schr\"{o}dinger equation.

\subsubsection{The potential case}

Next we study the case where the Lagrangian includes the potential 
$V$, which we express as 
\begin{equation}
V=\sum_{n=2}  b_n q^n .
\end{equation}
In this case since each component of the Taylor-expanded $L$ in Eq.(\ref{taylor_L}) 
is given by 
\begin{eqnarray}
&&L|_{\dot{q}=0} = -V=  - \sum_{n=2}  b_n q^n , \\
&&\frac{\partial L}{\partial \dot{q}}\bigg|_{\dot{q}=0}  = m\dot{q}|_{\dot{q}=0}  =0 , \\
&&\frac{\partial^2 L}{\partial \dot{q}^2 }\bigg|_{\dot{q}=0} = m, 
\end{eqnarray}
Eq.(\ref{time_dev_qbraAket2}) is calculated as 
\begin{eqnarray}
&&{}_m \langle_{new}~ q_{t+dt} | \psi(t+dt) \rangle|_\xi \nonumber \\ 
&=&
\frac{C_A}{\alpha}  ~{}_m \langle \text{anti} ~\xi | \psi(t) \rangle 
\int_C dq_t 
\exp\left[ -\frac{i}{\hbar} \sum_{n=2}  b_n q_t^n dt 
- \frac{i}{\hbar}  \frac{m}{dt} ( q_{t+dt} -\xi ) q_t 
+ \frac{i m}{2\hbar} \frac{q_{t+dt}^2}{dt}    -\frac{im}{2\hbar \Delta t} \xi^2
\right] \nonumber \\
&=&
\frac{C_A}{\alpha}  ~{}_m \langle \text{anti} ~\xi | \psi(t) \rangle 
\int_C \exp\left[ \frac{i}{\hbar}  (\xi - q_{t+dt} ) \frac{m}{dt} q_t \right] d q_t 
\exp\left[  \frac{i m}{2\hbar dt}  (q_{t+dt}^2 - \xi^2 ) \right] \nonumber \\
&&
- \frac{i}{\hbar} dt 
\frac{C_A}{\alpha}  ~{}_m \langle \text{anti} ~\xi | \psi(t) \rangle 
\int_C \sum_{n=2}  b_n q_t^n  
\exp\left[ \frac{i}{\hbar}  (\xi - q_{t+dt} ) \frac{m}{dt} q_t \right] d q_t 
\exp\left[  \frac{i m}{2\hbar dt}  (q_{t+dt}^2 - \xi^2 ) \right] . \nonumber \\
\label{time_dev_qbraAket_k=0general_a}
\end{eqnarray}
In the last expression we note that the first line is the same 
as Eq.(\ref{time_dev_qbraAket_k=0}), and for the second line 
we shall use the following relation,  
\begin{equation}
\int_C \exp\left[ \frac{i}{\hbar}  (\xi - q_{t+dt} ) q_t \right]
\frac{m}{dt} 
q_t^n d q_t = 
\left(  \frac{\hbar dt}{m} \right)^{n+1} 
(-i)^n 2\pi \frac{ \partial^n \delta_c(\xi - q_{t+dt} ) }{  \partial (\xi - q_{t+dt} )^n } .
\label{general_potential_kzero1}
\end{equation}
Then Eq.(\ref{time_dev_qbraAket_k=0general_a}) is expressed as 
\begin{eqnarray}
&&{}_m \langle_{new}~ q_{t+dt} | \psi(t+dt) \rangle|_\xi \nonumber \\ 
&=&
\frac{C_A}{\alpha}  ~{}_m \langle \text{anti} ~\xi | \psi(t) \rangle 
2\pi\hbar \delta_c(\xi - q_{t+dt}) \frac{dt} {m}
\exp\left[  \frac{i m}{2\hbar dt}  (q_{t+dt}^2 - \xi^2 ) \right]  \nonumber \\
&& - \frac{C_A}{\alpha} ~{}_m \langle \text{anti} ~\xi | \psi(t) \rangle \nonumber \\ 
&& \times
\sum_{n=2}  \left(  \frac{\hbar dt}{m} \right)^{n+1} 
(-i)^n 2\pi 
\frac{i dt}{\hbar}  b_n 
\exp\left[  \frac{i m}{2\hbar dt}  (q_{t+dt}^2 - \xi^2 ) \right]  
\frac{ \partial^n \delta_c(\xi - q_{t+dt} ) }{  \partial \xi^n }  . \nonumber \\
\label{time_dev_qbraAket_k=0general}
\end{eqnarray}
This result shows that ${}_m \langle_{new}~ q_{t+dt} | \psi(t+dt) \rangle|_\xi$ 
is equal to the linear combination of $\delta_c (q_{t+dt} -\xi)$ and its derivative. 
Therefore, as shown in the free case, 
only the component with $\xi=q_{t+dt}$ contributes to 
the value of ${}_m \langle_{new}~ q_{t+dt} | \psi(t+dt) \rangle$. 
We have thus obtained the momentum relation of Eq.(\ref{def_p_delLdelqdot}) 
in the potential case from the FPI point of view.

Next we calculate ${}_m \langle_{new}~ q_{t+dt} | \psi(t+dt) \rangle$ by integrating out 
Eq.(\ref{time_dev_qbraAket_k=0general}) with regard to $\xi$.
We have already seen that the integration of the first line of 
Eq.(\ref{time_dev_qbraAket_k=0general}) with regard to $\xi$ 
gives the last expression of Eq.(\ref{time_dev_qbraAket_k=0_2}). 
The integration of the second line of Eq.(\ref{time_dev_qbraAket_k=0general})  gives 
\begin{eqnarray}
&&\int_C d\xi \left[ {}_m \langle_{new}~ q_{t+dt} | \psi(t+dt) \rangle|_\xi \right]_{\text{potential part}} 
\nonumber \\ 
&=& 
- \frac{C_A}{\alpha} \int_C d\xi 
~{}_m \langle \text{anti} ~\xi | \psi(t) \rangle 
\sum_{n=2} 
\left(  \frac{\hbar dt}{m} \right)^{n+1} 
(-i)^n 2\pi 
\frac{i dt}{\hbar}  \nonumber \\
&&\times 
b_n 
\exp\left[  \frac{i m}{2\hbar dt}  (q_{t+dt}^2 - \xi^2 ) \right] 
\frac{ \partial^n \delta_c(\xi - q_{t+dt} ) }{  \partial \xi^n } \nonumber \\
&=&
-\frac{C_A}{\alpha} C_B^*   
\frac{i dt}{\hbar}   2\pi 
\sum_{n=2} i^n \left(  \frac{\hbar dt}{m} \right)^{n+1} b_n \nonumber \\
&&
\times 
\int_C dq_t \exp\left[  \frac{i m}{2\hbar dt}  (q_t^2 + q_{t+dt}^2 ) \right] 
~{}_m \langle_{new}~ q_t |  \psi(t) \rangle
\int_C d\xi \delta_c (\xi - q_{t+dt} ) 
\frac{ \partial^n  }{  \partial \xi^n } 
\exp\left[  \frac{-i m q_t}{\hbar dt}  \xi \right] \nonumber \\
&=&
- \frac{1}{\alpha}\frac{i dt}{\hbar}    \sum_{n=2}  b_n 
\int_C dq_t q_t^n  \exp\left[  \frac{i m}{2\hbar dt}  (q_t - q_{t+dt} )^2 \right] 
~{}_m \langle_{new}~ q_t |  \psi(t) \rangle \nonumber \\
&\simeq&
- \frac{i dt}{\hbar} 
~{}_m \langle_{new}~  q_{t+dt} | V(\hat{q}_{new}) | \psi(t) \rangle , \label{kzero_secondline}
\end{eqnarray}
where in the last relation we have used the following relation,  
\begin{equation}
\int_C q_t^n \exp \left[  -A( q_t - q_c)^2 \right] dq_t \simeq 
\sqrt{ \frac{\pi}{A}} q_c^n , \label{qnexpqq-qc2} 
\end{equation}
which stands for large positive $\text{Re} A$.

From Eqs.(\ref{time_dev_qbraAket_k=0_2}), (\ref{time_dev_qbraAket_k=0general}) and (\ref{kzero_secondline}) 
we obtain the following result,  
\begin{eqnarray}
{}_m \langle_{new}~ q_{t+dt} | \psi(t+dt) \rangle 
&=&\int_C d\xi ~{}_m \langle_{new}~ q_{t+dt} | \psi(t+dt) \rangle|_\xi    \nonumber \\ 
&\simeq& 
~{}_m \langle_{new}~  q_{t+dt} |  \exp \left[ -\frac{i}{\hbar} \hat{H} dt \right] |\psi(t) \rangle , 
\label{time_dev_qbraAket_k=0_3}
\end{eqnarray}
where $\hat{H}$ is given by 
\begin{equation}
\hat{H}  = \hat{H}_0 +  V(\hat{q}_{new}) . 
\end{equation}
Therefore also in the potential case we have obtained the Hamiltonian $H$ 
from the Lagrangian $L$ via FPI.  
In addition, Eq.(\ref{time_dev_qbraAket_k=0_3}) is reduced to 
$|\psi(t+dt) \rangle = e^{-\frac{i}{\hbar} \hat{H} dt}|\psi(t) \rangle$. 
Thus we have derived the Schr\"{o}dinger equation.

In this section we have derived the momentum relation of Eq.(\ref{def_p_delLdelqdot}), 
Hamiltonian, and Schr\"{o}dinger equation 
via FPI starting from the Lagrangian. 
Such a derivation of the Schr\"{o}dinger equation 
is well known in the RAT~\cite{FPIbook}. 
The analysis of this section has identified the definitions of the momentum and 
Hamiltonian in the CAT, and we have seen that they have the same forms as in the RAT. 
In the next section we make an analysis in the opposite way. 
We shall attempt to derive the Lagrangian starting from the Hamiltonian 
in FPI. In the analysis we shall find again the momentum relation 
of Eq.(\ref{def_p_delLdelqdot}).

\section{The derivation of the Lagrangian and the momentum relation}\label{derivation_lagrangian}

In this section  starting from the Hamiltonian which we 
obtained in the previous section, we shall derive a Lagrangian 
and see that it coincides with the Lagrangian from which we started in the previous section. 
This study can be performed as done in the RAT\cite{swanson,FPIbook}, 
except for the use of the newly introduced devices such as a modified bra. 
In addition we shall obtain the momentum relation of Eq.(\ref{def_p_delLdelqdot}) 
in a different way from the previous section.

We start from the following Hamiltonian 
\begin{equation}
\hat{H}(\hat{p}_{new}, \hat{q}_{new}) = \frac{\hat{p}_{new}^2 }{2m} + V(\hat{q}_{new} ) , 
\end{equation}
which we obtained in the previous section. 
The transition amplitude from an initial state 
$| i \rangle$ at time $t_i$ to a final state $| f \rangle$ at time $t_f$ is written as 
\begin{eqnarray}
&& \langle f |   e^{ - \frac{i}{\hbar} \hat{H} (t_f - t_i ) }  | i \rangle \nonumber \\
&=& \int_C dq_1 \cdots dq_{N}~  
\langle f | q_N \rangle_{new} 
~{}_m \langle_{new}~ q_N |  e^{ - \frac{i}{\hbar} \hat{H} \Delta t } | q_{N-1} \rangle_{new}~ 
~{}_m \langle_{new}~ q_{N-1} | \cdots | q_2 \rangle_{new}~ \nonumber \\
&&
\times
~{}_m \langle_{new}~ q_2 | e^{ - \frac{i}{\hbar} \hat{H} \Delta t } | q_1 \rangle_{new}~ 
~{}_m \langle_{new}~ q_1 | i \rangle ,  \label{braqbqaket}
\end{eqnarray}
where we have divided the time interval $t_f -t_i$ into $N-1$ pieces whose interval is 
\begin{equation}
\Delta t = \frac{ t_f - t_i } {N-1} .  
\end{equation}
Using this $\Delta t$, we define $\dot{q}_j$ as 
\begin{equation}
\dot{q_j} \equiv \frac{ q_{j+1} - q_j }{ \Delta t }  .  \label{def_dotq_j}
\end{equation}

To evaluate Eq.(\ref{braqbqaket}) we calculate 
${}_m \langle_{new}~ q_{j+1} | e^{ - \frac{i}{\hbar} \hat{H} \Delta t } | q_j \rangle_{new}$ 
considering the coordinate ordering implicitly as in the RAT as follows, 
\begin{eqnarray}
&&{}_m \langle_{new}~ q_{j+1} | e^{- \frac{i}{\hbar}  \hat{H}(\hat{p}_{new} , \hat{q}_{new}) \Delta t  } 
| q_{j} \rangle_{new}   \nonumber \\
&=&  \int_C dp_j 
e^{- \frac{i}{\hbar}   H(p_j, q_j)  \Delta t  } ~{}_m \langle_{new}~ q_{j+1} | p_j \rangle_{new} 
~{}_m \langle_{new}~ p_j | q_{j} \rangle_{new}~ \nonumber \\
&=& 
\int_C \frac{dp_j}{2\pi\hbar} 
\exp\left[  \frac{i}{\hbar} \Delta t L(p_j, q_j, \dot{q}_{j}) \right] , \label{braqj+1qjket}
\end{eqnarray}
where we have used Eq.(\ref{def_dotq_j}) and the following relations,  
\begin{eqnarray}
&&\int_C | p_j \rangle_{new} ~{}_m \langle_{new}~ p_j | dp_j = 1 , \\
&&{}_m \langle_{new}~ q_{j+1} |  p_j \rangle_{new}~ 
=\frac{1}{ \sqrt{ 2\pi\hbar} } \exp\left[ \frac{i}{\hbar} p_j q_{j+1}  \right]  , \\
&&{}_m \langle_{new}~ p_j |  q_{j} \rangle_{new}~ 
=\frac{1}{ \sqrt{ 2\pi\hbar} } \exp\left[ -\frac{i}{\hbar} p_j q_{j}  \right] . 
\end{eqnarray}
In the last line of Eq.(\ref{braqj+1qjket}) $L(p_j, q_j, \dot{q}_{j})$ is given by 
\begin{eqnarray}
L(p_j, q_j, \dot{q}_{j}) &=&p_j \dot{q}_{j} - H(p_j, q_j) \nonumber \\
&=& p_j \dot{q}_{j} - \frac{p_j^2}{2m}  - V(q_j) .
\end{eqnarray}

By substituting Eq.(\ref{braqj+1qjket}) for Eq.(\ref{braqbqaket}) 
$\langle f |   e^{ - \frac{i}{\hbar} \hat{H} (t_f - t_i ) }  | i \rangle$ is calculated as 
\begin{eqnarray}
&&\langle f |   e^{ - \frac{i}{\hbar} \hat{H} (t_f - t_i ) }  | i \rangle \nonumber \\
&=&
\int_C \frac{dp_1}{2\pi\hbar} \cdots \frac{dp_{N-1}}{2\pi\hbar} dq_1 \cdots dq_{N}~ 
\langle f | q_N \rangle_{new}~  ~{}_m \langle_{new}~ q_1 | i \rangle
\exp\left[  \frac{i}{\hbar} \sum_{j=1}^{N-1} \Delta t  L(p_j, q_j, \dot{q}_j) \right] \nonumber \\
&=& \int_C  {\cal D} p {\cal D} q~ \psi_f(q_f)^{*_{q_f}} \psi_i(q_i)
\exp\left[  \frac{i}{\hbar} \int_{t_i}^{t_f} dt L(p, q, \dot{q})  \right] ,
\end{eqnarray}
where in the second equality we have introduced $q_i=q_1$ and $q_f=q_N$. 
Also, we have taken the large $N$ limit  
and introduced ${\cal D} q$ and ${\cal D} p$ by 
\begin{eqnarray}
&&{\cal D} q = \lim_{N \rightarrow \infty}  
dq_1 \cdots dq_{N} , \\
&&{\cal D} p = \lim_{N \rightarrow \infty}  
\frac{dp_1}{2\pi\hbar} \cdots \frac{dp_{N-1}}{2\pi\hbar} . 
\end{eqnarray}
The integral with regard to $p_j$ is performed as follows,
\begin{eqnarray}
&&\int_C \frac{dp_j}{2\pi\hbar}  \exp\left[ \frac{i}{\hbar} \Delta t L(p_j, q_j, \dot{q}_j) \right] \nonumber \\
&=& 
\int_C \frac{dp_j}{2\pi\hbar} 
\exp\left[ \frac{i}{\hbar} \Delta t 
\left\{ -\frac{1}{2m}  \left( p_j - m \dot{q}_j \right)^2 
+ \frac{ 1}{2} m \dot{q_j}^2-V(q_j) \right\} \right] \nonumber \\
&=&\sqrt{ \frac{m}{2\pi i \hbar \Delta t} } \exp\left[  \frac{i}{\hbar} \Delta t L(\dot{q}_j , q_j)  \right] , 
\end{eqnarray}
where in the last equality based on the condition (\ref{m_I_positive_cond}) 
we have performed the Gaussian integral around the saddle point for $p_j$, 
\begin{equation}
p_j=m \dot{q_j} , \label{pjmqj}
\end{equation}
and in the last expression $L(\dot{q_j}, q_j)$ is given by 
\begin{equation}
L(\dot{q_j}, q_j) = \frac{1}{2} m \dot{q_j}^2 - V(q_j) . \label{lagrangian_dotq_q}
\end{equation}
From Eqs.(\ref{pjmqj}) and (\ref{lagrangian_dotq_q}) we obtain the momentum relation (\ref{def_p_delLdelqdot}).  
Thus we have obtained 
\begin{eqnarray}
\langle f |   e^{ - \frac{i}{\hbar} \hat{H} (t_f - t_i ) }  | i \rangle
&=&
\int_C \bar{\cal D} q~ \psi_f(q_f)^{*_{q_f}} \psi_i(q_i)
\exp\left[  \frac{i}{\hbar} \int_{t_i}^{t_f} dt L( \dot{q}, q) \right] , 
\end{eqnarray}
where 
\begin{equation}
\bar{\cal D} q = \lim_{N \rightarrow \infty}  \left( \frac{m}{2\pi i \hbar \epsilon} \right)^{\frac{N-1}{2}} dq_1 \cdots dq_{N}  . 
\end{equation}

In this section we have derived the Lagrangian (\ref{lagrangian_dotq_q}) 
starting from the Hamiltonian,  
and have seen that the Lagrangian has the same form as that in the previous section. 
We have thus confirmed that the Hamiltonian in the CAT has the same form as that in 
the RAT, and justified 
the classical construction of the Hamiltonian $H = p \dot{q} - L$ via FPI. 
Also, we have obtained the momentum relation (\ref{def_p_delLdelqdot}) 
via the saddle point for $p$. 
This is the second derivation of the momentum relation, 
which is different from that in the previous section. 
We present the third derivation of the momentum relation in appendix~\ref{via_saddle_q}.

Before ending this section, we discuss another possibility to define our Hamiltonian. 
The Hamiltonian $H$ we have derived 
is not hermitian, so we can decompose it as 
\begin{equation}
H=H_h + H_a, 
\end{equation}
where $H_h$ and $H_a$ are the hermitian and anti-hermitian part of $H$ respectively, and 
they are defined by 
\begin{eqnarray}
&&H_h=\frac{H+H^\dag}{2} , \label{H_{h}} \\ 
&&H_a=\frac{H-H^\dag}{2} . \label{H_{a}}
\end{eqnarray}
If we adopt $H_h$ as our Hamiltonian, then we can define a unitary time-development 
operator $U=e^{-\frac{i}{\hbar} H_h t}$ satisfying $U^\dag U =1$, 
while we have $H_h \neq p \dot q -L$. 
If we define a hermitian Lagrangian $L_h=\frac{L+L^\dag}{2}$, 
we obtain the relation $H_h = p \dot q -L_h$, which has a similar form as a usual one. 
But using $L_h$ means that we lose the complexity of the action, 
so this does not match our motivation to study the effect of the 
complex action. 
In addition, if $H$ gives time-development, $[H_h, H]\neq 0$. 
Hence the option to use both of $H$ and $H_h$ does not seem to work well. 
Therefore we adopt the Hamiltonian $H$ as our Hamiltonian. 
In other words we prefer the relation $H = p \dot q -L$ to 
the hermiticity of the Hamiltonian.

\section{Summary and outlook}

In the complex action theory (CAT) even a coordinate $q$ and a momentum $p$ 
can be complex in general. 
In Ref.~\cite{Nagao:2011za}, to handle the complex $q$ and $p$, 
we have proposed the replacement of hermitian operators of coordinate and momentum 
$\hat{q}$ and $\hat{p}$ and their eigenstates $\langle q |$ and $\langle p |$ 
with non-hermitian operators $\hat{q}_{new}$ and 
$\hat{p}_{new}$, and ${}_m\langle_{new}~ q |$ and ${}_m\langle_{new}~ p |$ with complex eigenvalues $q$ and $p$.
We have introduced a philosophy of keeping the analyticity 
in parameter variables of Feynman path integral (FPI), 
and defined modified bras ${}_m \langle ~|$ 
and ${}_{\{ \}} \langle ~|$, a modified complex conjugate $*_{ \{ \} }$, 
and a modified hermitian conjugate $\dag_{\{\}}$ to realize the philosophy, 
by means of which we have explicitly constructed $\hat{q}_{new}$,  
$\hat{p}_{new}$, and the hermitian conjugates of their eigenstates 
$| q \rangle_{new}$ and $| p \rangle_{new}$ by utilizing the coherent states of 
harmonic oscillators.
Here $\{\}$ denotes a set of parameters which we keep the analyticity in. 
In this paper, based on the complex coordinate formalism of Ref.~\cite{Nagao:2011za}, 
we have explicitly examined how the momentum and Hamiltonian are defined 
in the CAT from the FPI point of view.

In section 2 we have briefly reviewed the complex coordinate 
formalism of Ref.~\cite{Nagao:2011za}. 
First we have explained that the delta function can be used also for complex parameters 
when it satisfies some condition, and introduced the new devices to realize 
the philosophy of keeping the analyticity in parameter variables of FPI. 
The philosophy allows us to deform the path of FPI. 
Then we have briefly shown the construction of non-hermitian operators $\hat{q}_{new}$ and 
$\hat{p}_{new}$, and  the hermitian conjugates of their eigenstates 
$| q \rangle_{new}$ and $| p \rangle_{new}$ with complex eigenvalues $q$ and $p$ 
by utilizing the coherent states of harmonic oscillators.  
In section~\ref{derivation_hamiltonian} based on the philosophy and using the new devices 
explained in section~\ref{fundamental} 
we have described in FPI with a Lagrangian the time development of a $\xi$-parametrized wave function, 
which is a solution to an eigenvalue problem of a momentum operator. 
Solving the eigenvalue problem and considering the time-development of 
the wave function solution in FPI we have derived the momentum relation 
$p=\frac{\partial L}{\partial \dot{q}}$. 
Furthermore we have also studied the form of the Hamiltonian in the CAT. 
We have derived the Hamiltonian and Schr\"{o}dinger equation starting from the Lagrangian in FPI, 
first in a free theory case, and then in a potential case. 
In section~\ref{derivation_lagrangian}, oppositely 
starting from the Hamiltonian, we have derived the Lagrangian in a potential case via FPI, 
and also the momentum relation again via the saddle point for $p$.
These explicit studies have confirmed that both of the momentum and Hamiltonian in the CAT 
have the same forms as those in the real action theory (RAT). 
In appendix~\ref{via_saddle_q} we have explained the third derivation of the momentum relation 
via the saddle point for $q$.

Now the momentum and Hamiltonian in the CAT have been identified clearly, 
based on the complex coordinate formalism of Ref.~\cite{Nagao:2011za}. 
As a next step what should we study to develop the CAT? 
One direction is to study a future-included theory, 
that is to say, a theory including not only a past time 
but also a future time as an integration interval of time, 
which we have not considered in this paper, in the complex coordinate formalism. 
It is interesting to study such a future-included theory by considering 
a kind of wave function of universe including the future information, which was 
discussed in Ref.~\cite{Nielsen:2007mj}. 
In addition, it is also important to study an expectation value in the CAT. 
We shall attempt to discuss how the expectation value is defined in the CAT generally, 
and also clarify its relation to the modified hermiticity. 
Furthermore, it is also intriguing to investigate in detail 
the possible misestimation of the past state 
by extrapolating back in time with the hermitian Hamiltonian. 
As pointed out in Ref.~\cite{Nagao:2010xu}, the misestimation by 
a historian living in the late time can occur in some fundamental theory which 
is described by a non-hermitian Hamiltonian. 
We will study them and report the progress in the future.


\section*{Acknowledgments}

This work was partially funded by Danish Natural Science Research Council (FNU, Denmark), 
and the work of one of the authors (K.N.) was supported in part by 
Grant-in-Aid for Scientific Research (Nos.18740127 and 21740157) 
from the Ministry of Education, Culture, Sports, Science and Technology (MEXT, Japan). 
K.N. would like to thank all the members and visitors of NBI for their kind hospitality.

\appendix

\section{The third derivation of the momentum relation}\label{via_saddle_q}

In this section we give the third derivation of the momentum relation of 
Eq.(\ref{def_p_delLdelqdot}). 
Taking a momentum eigenstate 
\begin{equation}
\Psi_{p}(q)=e^{\frac{i}{\hbar} p q } , \label{momeneigen} 
\end{equation}
where $q$ and $p$ are in general complex, 
we study the following path integral for an infinitesimal time from $t$ to $t+dt$, 
\begin{eqnarray}
&&\int_C \Psi_{p'} (q_{t+dt})^{*_{q_{t+dt}}}  e^{\frac{i}{\hbar} L(q, \dot{q}) dt} \Psi_{p}(q_t) 
dq_t dq_{t+dt} \nonumber \\
&=&
\int_C e^{-\frac{i}{\hbar} (p' q_{t+dt} -p q_t -L(q, \dot{q}) dt ) } dq_t dq_{t+dt}, 
\end{eqnarray}
where $\dot{q}$ is given by Eq.(\ref{dot_q}). 
Then performing the Taylor-expansion of the Lagrangian $L(q,\dot{q})$ 
around $\dot{q}=0$ as in Eq.(\ref{taylor_L}), we have 
\begin{eqnarray}
&&\int_C e^{-\frac{i}{\hbar} (p' q_{t+dt} -p q_t -Ldt ) } dq_t dq_{t+dt} \nonumber \\
&=&
\int_C 
\exp\left[  
\frac{i}{2\hbar}  \frac{\partial^2 L}{\partial \dot{q}^2}\bigg|_{\dot{q}=0}  \frac{1}{dt}
\left( q_t - 
\left\{  q_{t+dt} 
+ \frac{dt}{ \frac{\partial^2 L}{\partial \dot{q}^2 }\big|_{\dot{q}=0}  }
\left( \frac{\partial L}{\partial \dot{q}}\bigg|_{\dot{q}=0} - p \right)
\right\} 
\right)^2
\right. \nonumber \\
&&\left.
+ \frac{i}{\hbar} q_{t+dt} (p -p') + \frac{i}{\hbar} dt L
- \frac{i}{2\hbar}  \frac{dt}{ \frac{\partial^2 L}{\partial \dot{q}^2 } \big|_{\dot{q}=0} }
\left( \frac{\partial L}{\partial \dot{q}}\bigg|_{\dot{q}=0} - p \right)^2
\right]
dq_t dq_{t+dt} , \nonumber \\
\end{eqnarray} 
where we note that the Gaussian function is very narrow like the delta function 
since $dt$ is assumed to be a very small quantity. 
The saddle point $q_{t, saddle}$ is given by
\begin{equation}
q_{t, saddle} =q_{t+dt} 
+ \frac{dt}{ \frac{\partial^2 L}{\partial \dot{q}^2 }\big|_{\dot{q}=0}  } 
\left( \frac{\partial L}{\partial \dot{q}}\bigg|_{\dot{q}=0}  - p \right) ,
\end{equation}
and we have the relation
\begin{eqnarray}
\dot{q}_{t, saddle} &=&
\frac{q_{t+dt} - q_{t, saddle} }{dt} \nonumber \\
&=&-\frac{1}{ \frac{\partial^2 L}{\partial \dot{q}^2 }\big|_{\dot{q}=0} }
\left( \frac{\partial L}{\partial \dot{q}}\bigg|_{\dot{q}=0} - p \right)  \nonumber \\
&=&-\frac{1}{ \frac{\partial^2 L}{\partial \dot{q}^2 }\big|_{\dot{q}=0} }
\left( \frac{\partial L}{\partial \dot{q}}
 - p \right) + \dot{q}_{t, saddle} , \label{qsaddle_momentum}
\end{eqnarray}
where in the last equality 
we have used the Taylor-expansion of $\frac{\partial L}{\partial \dot{q}} (q, \dot{q})$,  
\begin{equation}
\frac{\partial L}{\partial \dot{q}} (q, \dot{q}) = \frac{\partial L}{\partial \dot{q}}\bigg|_{\dot{q}=0}  
+ 
\dot{q} \frac{\partial^2 L}{\partial \dot{q}^2 }\bigg|_{\dot{q}=0} ~.
\end{equation} 
From Eq.(\ref{qsaddle_momentum}) we have thus derived the momentum relation of Eq.(\ref{def_p_delLdelqdot}) 
by seeing the dominant saddle point in $q$.

\end{document}